\documentclass[3p]{elsarticle}

\usepackage{aas_macros}
\usepackage{lineno,hyperref}
\hypersetup{
  colorlinks,
  citecolor=Blue,
  linkcolor=Red,
  urlcolor=Blue}

\usepackage[table]{xcolor}
\usepackage{cancel}

\usepackage{cleveref}

\usepackage{caption}
\usepackage{subcaption}

\journal{Icarus}

\biboptions{authoryear}

\renewcommand{\deg}{^\circ}

\newcommand{\arcsec}{^{\prime\prime}}

\newcommand{\au}{\,\mathrm{\au}}
\newcommand{\km}{\,\mathrm{km}}
\newcommand{\meter}{\,\mathrm{m}}
\newcommand{\cm}{\,\mathrm{cm}}
\newcommand{\mm}{\,\mathrm{mm}}
\newcommand{\um}{\,\mu \mathrm{m}}

\newcommand{\hour}{\,\mathrm{h}}

\newcommand{\magnitude}{\,\mathrm{mag}}
\newcommand{\K}{\,\mathrm{K}}

\newcommand{\Pa}{\,\mathrm{Pa}}
\newcommand{\kg}{\,\mathrm{kg}}

\newcommand{\W}{\,\mathrm{W}}

\newcommand{\tiu}{\,\mathrm{J} \mathrm{K}^{-1} \mathrm{m}^{-2}\mathrm{s}^{-1/2}}

\begin{document}

\begin{frontmatter}

\title{Evidence of surface heterogeneity on active asteroid (3200) Phaethon}

\author{Eric MacLennan}
\address{Department of Physics, PO Box 64, 00014 University of Helsinki, Finland}

\author{Sean Marshall}
\address{Arecibo Observatory, University of Central Florida, HC-3 Box 53995, Arecibo, PR 00612, U.S.A.}

\author{Mikael Granvik}
\address{Department of Physics, PO Box 64, 00014 University of Helsinki, Finland}
\address{Asteroid Engineering Laboratory, Lule\r{a} University of Technology, Box 848, SE-98128 Kiruna, Sweden}





\begin{abstract}
Thermal infrared emission and thermophysical modeling techniques are powerful tools in deciphering the surface properties of asteroids. The near-Earth asteroid (3200) Phaethon is an active asteroid with a very small perihelion distance and is likely the source of the Geminid meteor shower. Using a thermophysical model with a non-convex shape of Phaethon we interpret thermal infrared observations that span ten distinct sightings. The results yield an effective diameter of $5.4 \pm 0.1 \km$ and independent thermal inertia estimates for each sighting. We find that the thermal inertia varies across each of these sightings in a way that is stronger than the theoretical temperature-dependent expectation from radiative heat transfer within the regolith. Thus, we test whether the variation in thermal inertia can be explained by the presence of a regolith layer over bedrock, or by a spatially heterogeneous scenario. We find that a model in which Phaethon's hemispheres have distinctly different thermophysical properties can sufficiently explain the thermal inertias determined herein. In particular, we find that a boundary is located between latitudes $-30\deg$ and $+10\deg$ that separates two regions: a fine-grained southern latitudes and a northern hemisphere that is dominated by coarse-grained regolith and/or a high coverage of porous boulders. We discuss the implications related to Phaethon's activity, potential association with 2005~UD, and the upcoming DESTINY$^+$ mission.

\end{abstract}


\end{frontmatter}


\section{Introduction}

Asteroid surface properties can be inferred from the interpretation of thermal inertia, $\Gamma$, which is an indicator of resistance to temperature changes \citep[][and references therein]{Delbo_etal15}. Thermal inertia is defined as $\Gamma = \sqrt{k \rho c_s}$, where $k$ (thermal conductivity), $\rho$ (mass density), and $c_s$ (specific heat capacity) are effective bulk properties of the near-surface. Both the grain size and porosity of the regolith are major factors that influence heat transport \citep[e.g.,][]{Wood20}, specifically the conduction through contacts between the regolith grains and the radiative transfer within the pores between them \citep{Watson64,Keihm_etal73,Gundlach&Blum13,Sakatani_etal17,Ryan_etal20}. Generally speaking, for atmosphereless bodies such as asteroids, the effective thermal conductivity can vary by several orders of magnitude with changes in the grain size \citep{Presley+Christensen97,Presley+Craddock06}. Investigations of asteroid surfaces using thermal inertias and so-called thermally characteristic grain sizes are particularly useful when investigating asteroid surface processes and in comparing individual objects to the broad population \citep{MacLennan+Emery21,MacLennan+Emery22}.

The near-Earth asteroid (NEA) (3200) Phaethon is one of the few asteroids linked to a meteor stream \citep{Williams+Wu1993,Kasuga+Jewitt19}. The orbital similarity to Geminid stream and observed dust activity during its perihelion passage \citep{Jewitt_etal2013,Hui&Li2016} indicates a link to the annual Geminid meteor shower \citep{Whipple83,Gustafson1989,Ryabova_etal2019}.  In less than a decade the JAXA DESTINY$^+$\footnote{\url{https://destiny.isas.jaxa.jp/science/}} mission will conduct in-situ flyby observations of Phaethon's dust environment \citep{Arai_etal2018,Arai_etal2019}. Its reflectance spectrum indicates a link to the carbonaceous chondrite meteorites \citep{Licandro_etal2007,Clark_etal10,deLeon_etal12,Kareta_etal18,Kareta_etal21}. However, there is no evidence of hydration expected from the presence of, e.g., phyllosilicates \citep{Takir_etal20}; which is most likely a result of intense heating during its perihelion passage \citep{MacLennan_etal21}.

The thermal inertia of Phaethon has been estimated to be $600 \pm 200 \tiu$ by \citet{Hanus_etal2016} and $880^{+580}_{-330} \tiu$ by \citet{Masiero_etal2019}. While \citet{Hanus_etal2016} used a convex shape model derived from lightcurve inversion techniques, \citet{Masiero_etal2019} assumed a spherical shape. While these thermal inertia estimates are consistent with one another, the effective diameter ($D_\mathit{eff}$--the diameter of an equal-volume sphere) estimated by \citet{Hanus_etal2016} and \citet{Masiero_etal2019} ($D_\mathit{eff} = 5.1 \pm 0.2 \km$ and $4.6^{+0.2}_{-0.3} \km$, respectively) are in slight disagreement when only the 1-$\sigma$ uncertainties are considered. It is not clear if the differences are due to shape assumptions, or the distinct observational datasets used in each respective investigation. Radar observations of Phaethon during its 2017 close approach to Earth indicate a spinning top shape with an effective diameter of $D_\mathit{eff} \sim 5.5 \km$ \citep{Taylor_etal2019}, which is consistent with the lengths of stellar occultation chords \citep{Ye_etal19,Herald_etal20}. The discrepancy between the smaller size estimates from thermal infrared and larger size derived from radar observations is yet to be reconciled.

In this work we reassess Phaethon's size and thermal inertia by using a detailed shape model derived from optical lightcurve and radar observations. This shape model is qualitatively similar to the soon-to-be published version, but differs in some respects: slightly different dimensions, and minor variations in some small surface features. After constraining the size using a subset of the observations, we estimate the thermal inertia for each epoch individually. Our TPM results are then analyzed in the context of the heliocentric distance of each observational epoch. In particular, we attempt to discern details about the regolith structure, and/or spatial variability in Phaethon's surface and speculate on how the findings relate to Phaethon's activity and potential relationship with 2005~UD.

\section{Data and Methods}\label{sub:tpm}

\subsection{Infrared Observations}

We collect and use all the publicly-available thermal infrared data for Phaethon. In total, there are 11 distinct observing epochs from 5 telescopes that represent a rich dataset acquired over a time span of nearly 36 years. Specifically, we use thermal observations from Infrared Astronomical Satellite (IRAS), United Kingdom Infrared Telescope (UKIRT), the Infrared Spectrograph (IRS) and Multiband Imaging Photometer for Spitzer (MIPS) instruments on the Spitzer Space Telescope (hereafter Spitzer), Akari satellite \citep{Usui_etal11}, and 6 sightings made by the Wide-Field Infrared Survey Explorer (WISE) mission and the follow-up NEOWISE survey \citep{Mainzer_etal2011,Mainzer_etal2014}. We describe below the telescopes and circumstances of each observation. A summary of all these data and observing geometries is given in \autoref{tab:thermobs}.

The IRAS (Infrared Astronomical Satellite) survey observed Phaethon simultaneously at 12, 25, 60, and 100$\um$ on 1983-Oct-11 \citep{Tedesco_etal02}. In total, six data points were collected in each of the three shorter wavelength bands, and we omit a single 100$\um$ observation in order to avoid the uncertainty of emissivity values at longer wavelengths \citep[e.g.,][]{Muller_etal14}.

Photometry at visible and infrared wavelengths were obtained using UKIRT on the nights of 1984-Dec-20,21 by \citet{Green_etal85}. The measurements acquired at $<4\um$ contain a significant amount of reflected visible light. To avoid possible contamination from reflected light, we only use the flux values at wavelengths from 4.7 to 19.2$\um$. Each photometric data point was acquired at an individual time and thus represent a different part of Phaethon's rotational phase. The Akari astronomical survey detected Phaethon at two wavelengths (9 and 18$\um$) on 2006-Sep-23 \citep{Usui_etal11}. The color-corrected Akari photometry are taken from the Asteroid catalog using Akari (AcuA) archive\footnote{\url{https://www.ir.isas.jaxa.jp/AKARI/Archive/Catalogues/Asteroid_Flux_V1/}}. 

The Spitzer Space Telescope observed Phaethon with the Infrared Spectrograph (IRS) on 2005-Jan-14 \citep{Hanus_etal2016} and with the Multiband Imaging Photometer \citep[MIPS;][]{Rieke_etal04} on 2007-Jan-01. We use the IRS observations---a high signal-to-noise spectrum that covers wavelengths in the range 5-$30\um$---analyzed by \citet{Hanus_etal2016}. This spectrum has an absolute calibration error of at least $5\%$ for each independent datapoint. The MIPS observation consists of one data point at 24$\um$ (effective wavelength of 23.675$\um$),  which has not been previously utilized. For the MIPS observations, we download the original image files via the Spitzer Heritage Archive\footnote{\url{https://www.sha.ipac.caltech.edu}} and construct an image mosaic using MOPEX software \citep{Makovoz_etal05}. Aperture photometry is then performed using a radius of $3.5\arcsec$ and a sky annulus defined by inner and outer radii of $20\arcsec$ and $32\arcsec$, respectively, for which we apply an aperture correction of 2.57\footnote{\url{https://www.irsa.ipac.caltech.edu/data/SPITZER/docs/mips/mipsinstrumenthandbook/}}. Finally we used a color correction of 0.956 for a $205\K$ blackbody \citep{Stansberry_etal07} to calculate a flux of 60.9 $\mu$Jy, for which we conservatively assign a flux uncertainty of 5\% \citep{Engelbracht_etal07}.

The WISE telescope observed Phaethon first during the cryogenic phase of the mission \citep{Mainzer_etal2011} in 2010 at a heliocentric distance of 2.317~au (near its aphelion distance of 2.403~au). At this distance, thermal emission dominated the W3 and W4 bands, approximately 12 and 22$\um$, respectively. The next five sightings occurred during the reactivated NEOWISE phase of the mission (2015-2019), during which only the W1 and W2-bands, approximately 3.4 and 4.6$\um$, respectively, were operational \citep{Mainzer_etal2014}. However, these NEOWISE observations were collected when Phaethon was closer to the Sun where the thermal contribution at these wavelengths was significant \citep{Masiero_etal2019}. Hereafter, we refer to the first sighting during the cryogenic mission as ``WISE'' and the latter five sightings as ``NEOWISE(X)'', where the ``X'' is the sighting number listed in \autoref{tab:thermobs}.

\begin{table}
\caption{Observation Summary of Phaethon}
    \centering
    \renewcommand{\arraystretch}{1.1}
    \begin{tabular}{ll|ccrrrr|lr}
        \hline \vspace{-0.2cm}
        UT Date & Telescope/Sighting & $r_\mathrm{au}$$^a$ & $\Delta_\mathrm{au}$$^b$ & $\alpha_\odot$$^c$ & $\phi_{ss}$$^d$ & $\phi_{so}$$^e$ & orb.$^f$ & $N_{obs}$$^g$ & $\lambda_{obs}$$^h$ ($\um$) \vspace{0.3cm} \\ \hline \\[-1,6em] \hline \vspace{-0.3cm} \\
        1983-Oct-11 & IRAS & 1.033 & 0.371 & $74.3\deg$ & $-3\deg$ & $45\deg$ & post-$q$ & 6, 6, 6 & 12, 25, 60 \\
        1984-Dec-20,21 & UKIRT & 1.131 & 0.246 & $48.3\deg$ & $24\deg$ & $0\deg$ & pre-$q$ & 12$^\star$ & 4.73 to 19.2 \\
        2005-Jan-14 & Spitzer-IRS & 1.125 & 0.502 & $-63.3\deg$ & $24\deg$ & $-16\deg$ & pre-$q$ & 344 & $5.1-37.8$ \\
        2006-Sep-23 & Akari & 1.122 & 0.510 & $-63.4\deg$ & $-2\deg$ & $61\deg$ & post-$q$ & 2$^\star$ & 9, 18 \\
        2007-Jan-01 & Spitzer-MIPS & 2.088 & 1.483 & $26.6\deg$ & $-7\deg$ & $4\deg$ & pre-$Q$ & 1 & 24 \\
        2010-Jan-07 & WISE &  2.317 & 2.077 & $25.1\deg$ & $10\deg$ & $-5\deg$ & pre-$Q$ & 1, 2, 2 & 3.4, 12, 22 \\
        2015-Jan-13 & NEOWISE(1) & 1.328 & 0.831 & $47.7\deg$ & $22\deg$ & $-8\deg$ & pre-$q$ & 4, 4 & 3.4, 4.6 \\
        2016-Oct-02 & NEOWISE(2) & 1.081 & 0.406 & $-67.8\deg$ & $-3\deg$ & $61\deg$ & post-$q$ & 3, 3 & 3.4, 4.6 \\
        2016-Dec-06,07 & NEOWISE(3) & 1.841 & 1.353 & $31.5\deg$ & $5\deg$ & $-8\deg$ & pre-$Q$ & 19, 19 & 3.4, 4.6 \\
        2017-Dec-17 & NEOWISE(4) & 1.007 & 0.069 & $-68.9\deg$ & $25\deg$ & $10\deg$ & pre-$q$ & 1 & 3.4 \\
        2019-Sep-07 & NEOWISE(5) & 1.401 & 0.975 & $-46.0\deg$ & $1\deg$ & $40\deg$ & post-$q$ & 6, 6 & 3.4, 4.6 \\ \hline
        \multicolumn{10}{l}{$^a$Heliocentric distance (au).}\\
        \multicolumn{10}{l}{$^b$Observer-centric distance (au).}\\
        \multicolumn{10}{l}{$^c$Solar phase angle. Negative and positive $\alpha_\odot$ values indicate, respectively, pre- and post-opposition.}\\
        \multicolumn{10}{l}{$^d$Sub-solar latitude.}\\
        \multicolumn{10}{l}{$^e$Sub-observer latitude.}\\
        \multicolumn{10}{l}{$^f$Orbital position relative to perihelion ($q$) or aphelion ($Q$).}\\
        \multicolumn{10}{l}{$^g$Number of observations in each band, sorted by effective wavelength.}\\
        \multicolumn{10}{l}{$^h$Approximate effective wavelength of each band.}\\
        \multicolumn{10}{l}{$^\star$Data for each band are acquired at independent times.}
    \end{tabular}
    \label{tab:thermobs}
\end{table}

\subsection{Shape Model Properties}\label{sub:shape}

The preliminary shape model used herein \citep{Marshall_etal21} was derived using lightcurve observations from 16 apparitions (from 1989 to 2019), and radar data from Arecibo Observatory (from 2007 and 2017) and from the Goldstone Deep Space Communications Complex (from 2017 only). Shape modeling was done using the SHAPE software \citep{Magri_etal_2007}. A publication describing the final shape model is in preparation. The modeled rotation period ($P_\mathit{rot} = 3.603955 \hour$) and spin axis orientation (ecliptic longitude and latitude of $\lambda = 316\deg$ and $\beta = -48.7\deg$, respectively) are very similar to the convex model of \citet{Hanus_etal2018}. The ratios of the Dynamically Equivalent Equal-Volume Ellipsoid (DEEVE) axes are $a/b = 1.08$ and $b/c = 1.20$. This shape model has a mean (volume-equivalent) diameter of $\sim5.38 \km$, based on constraints from radar observations. The occultation chords were not directly incorporated into this stage of modeling, but they suggest an average diameter about 4\% smaller than the best-fit radar size, consistent within the expected uncertainties. Note that we scale the size of the shape model in our TPM to independently derive a best-fit size from the thermal observations.

Phaethon's rotation phase is known from joint lightcurve and radar observations that span over most of the dates in the infrared dataset. This is particularly useful for the NEOWISE(2) and NEOWISE(4) data that consist of only a few datapoints. The exceptions are for the IRAS and UKIRT epochs, in which Phaethon's rotation phase in the 1980s would be accurately known if its rotation period was identical to the value observed from 1994-2019. However, as noted in \citet{Hanus_etal2016}, a single lightcurve from 1989 by \citet{Wisniewski_etal97} seems to have a different rotation phase\footnote{this dataset was subsequently rejected by \citet{Hanus_etal2016}}, and there are no other published lightcurve observations from before 1994 that might allow us to resolve this discrepancy. Phaethon's rotation period may have changed slightly (possibly due to activity near perihelion), or perhaps there is an unaccounted for error in the 1989 lightcurve. In any case, the uncertainty in rotation phase becomes too large when the information from lightcurves obtained after 1994 are extrapolated backwards by 10+ years. The respective rotation phases in the IRAS and UKIRT observations are thus left as free parameters and adjusted to minimize the $\chi^2$ for each set of input parameters.

\subsection{Thermophysical Modeling}

In this work we present and implement a TPM that incorporates a preliminary non-convex shape model of Phaethon (\autoref{sub:shape}) derived from radar and lightcurve observations. The TPM incorporates one-dimensional heat conduction, small-scale surface roughness, and global self-heating and shadowing effects; the latter two of which are relevant for non-convex shapes. Closely following the Advanced Thermophysical Model \citep{Rozitis+Green11}, we modify the TPM presented in \citet{MacLennan&Emery2019} to include features that account for the above effects and call it \texttt{shapeTPM}. The assumed thermophysical properties in \texttt{shapeTPM} are discussed below and are assumed to be the same across the surface, and constant with depth and temperature. We later explore in this work how these assumptions may not hold. However, in the context of thermal inertia determination for individual sightings, it is sufficient to assume these points.

The energy balance at the surface of an asteroid can be expressed as:
\begin{equation}\label{eq:energybal}
    E_\odot\cos(z)(1-A)(1-s) + E_\mathit{scat} + E_\mathit{rad} + k_\mathit{eff} \frac{\Delta T}{\Delta x}\Big|_\mathit{surf} - \varepsilon_B \sigma T^4 = 0,
\end{equation}
where $E_\odot = S_\odot / r_\mathrm{au}^{2}$ is the insolation at 1~au, $S_\odot = 1367\ \mathrm{W} \meter^{-2}$, scaled to the desired heliocentric distance, $r_\mathrm{au}$. The first term in \autoref{eq:energybal} is the amount of direct insolation absorbed by a surface with bolometric Bond albedo $A$ (hereafter Bond albedo) and solar incidence angle, $z$ (the angle between a surface facet’s local zenith and Sun-direction). The effective thermal conductivity, $k_{eff}$, and surface temperature gradient, $\Delta T/\Delta x|_\mathit{surf}$, determine the subsurface heat conduction. Radiated energy is calculated using the Stefan–Boltzmann constant, $\sigma_0$, bolometric emissivity $\varepsilon_B$, and surface temperature, $T$. The total energy budget includes scattered solar radiation ($E_\mathit{scat}$) and re-absorbed thermal radiation ($E_\mathit{rad}$, i.e., self-heating), both of which occur among the large-scale facets of the shape model and between small-scale surface roughness elements. The shadowing factor, $s$, indicates if the surface (shape facet or roughness element) is shadowed by another part of the surface ($s=1$) or not ($s=0$). For instance, shape facets that exist in topographical lows can be shadowed by other facets and, at a smaller scale, rough elements often block sunlight from reaching other rough elements.

In order to quantify the energy exchange for shape facets and rough elements, we must calculate their view factors, i.e., the amount of radiation that is exchanged between two given areas. The view factor ($f_{j \to i}$ from facet $j$ to $i$) calculation can be approached in different ways, such as the summation via sub-sampled areas of a shape facet \citep{Rozitis+Green11}. We choose to perform a line integral around shape facet edges \citep[e.g.,][]{Lagerros98}, which has been shown to provide improved accuracy and is computationally faster than double-area integration \citep{Walton02}, particularly between facet pairs with a shared edge. We use the formulation of \citet{Mitalas+Stephenson66}:
\begin{equation}\label{eq:viewana}
    f_{j \to i} = \frac{1}{2 \pi a_j} \sum^3_{p=1} \sum^3_{q=1} \cos{\Phi_{pq}}\sum^N \sum^N \ln(r_{ij}) \Delta v_j \Delta v_i.
\end{equation}
Here, view factor subscript denotes that the radiation is received by facet $i$ from facet $j$, which have areas $a_i$ and $a_j$, respectively. The $q$th and $p$th edges of facet $i$ and $j$ are divided into $N$ segments of length $\Delta v_i$ and $\Delta v_j$, respectively. The distance between the midpoints of each segment pair is $r_{ij}$ and the angle between the edges is given by $\Phi_{pq}$. Each possible combination of edge pairs is considered. The derivation of \autoref{eq:viewana} is detailed in \ref{appA} and the variables from \autoref{eq:viewana} are shown in \autoref{fig:viewfac}.

View factors are used to compute the amount of reflected solar energy prior to the first time step. When only single scattering is considered, then the total amount of reflected insolation on facet $i$ from any other facets in view ($j$) is \citep{Rozitis+Green11}:
\begin{equation}
    E^\mathit{smooth}_{i,\mathit{scat}} =  A E_{\odot} \sum_{j\neq i} f_{j \to i} (1-s_j) \cos(z_j).
\end{equation}
Similarly, the total amount of thermal radiation received from other facets is:
\begin{equation}
    E^\mathit{smooth}_{i,\mathit{rad}} = \varepsilon_B \sigma_0 (1-A_{th}) \sum_{j\neq i} f_{j \to i} T^4_j
\end{equation}
and is dependent on facet surface temperatures and their Bond albedo at thermal wavelengths \citep[$A_{th} \sim 0$;][]{Rozitis+Green11}. The scattered solar energy received for each facet is computed before a TPM run as all the factors are independent of adjacent time steps. However, because temperatures are not known {\it a priori}, the self-heating terms must be solved for at each time step. The assumption of single scattering has been shown to be a sufficient for low Bond albedos \citep{Rozitis+Green11}, and because Phaethon's Bond albedo has been independently estimated to be $A\approx 0.048$ \citep{Hanus_etal2018} and $A\sim 0.05$ \citep{Shinnaka_etal18,Devogele_etal18} we claim that it is an appropriate approximation for our study.

As in \citet{MacLennan&Emery2019}, the modeling of small-scale surface roughness (i.e. on the order of the thermal skin depth) is approximated by using spherical section craters \citep{Hansen77,Emery_etal98}. The degree of roughness is determined by the opening angle ($\gamma$) of the spherical crater, measured from the center-line of the crater to the edge\footnote{A hemispherical crater has a $\gamma = 90\deg$, for example.}, and the fraction of surface area that is covered by craters ($f_R$). Independently changing these two parameters will alter the overall surface roughness, but contribute to mean surface slope, $\bar{\theta}$, via the following \citep{Lagerros96}:
\begin{equation}
    \tan{\bar{\theta}} = \frac{2f_R}{\pi}\frac{\sin({\gamma})-\ln[1+\sin({\gamma})]+\ln\cos({\gamma})}{\cos({\gamma})-1}.
\end{equation}
For small $\bar{\theta}$, different combinations of $\gamma$ and $f_R$ yield similar thermal emission profiles \citep{Emery_etal98}, however this is not true when either roughness parameter is large and is different for other roughness representations \citep[e.g., fractal surface;][]{Davidsson_etal15}. We pair crater opening angles with crater coverage fraction to define a set of three roughness values (in addition to a smooth surface) that we use for TPM fitting. The parameters for low, medium, and high default roughness surfaces are given in \autoref{tab:roughness}.

\begin{table}[h]
    \caption{Default TPM surface roughness values}\label{tab:roughness}
    \centering
    \begin{tabular}{lccc}
        \hline \vspace{-0.3cm} \\
        roughness & $\gamma$$^a$ & $f_R$$^b$ & $\bar{\theta}$$^c$ \vspace{0.15cm} \\ \hline \\[-1,4em] \hline \vspace{-0.3cm} \\
        smooth & | & | & 0$\deg$ \\
        low & $45\deg$ & 0.5 & $11\deg$ \\
        medium & $68\deg$ & 0.8 & $30\deg$ \\
        high & $90\deg$ & 1.0 & $58\deg$ \\ \hline
        \multicolumn{4}{l}{$^a$Crater opening angle.}\\
        \multicolumn{4}{l}{$^b$Fraction of crater coverage.}\\
        \multicolumn{4}{l}{$^c$Mean surface slope.}\\
    \end{tabular}
\end{table}

The incident energy for crater elements will differ from that of a smooth surface, as seen in the energy budget calculation. First, we calculate the scattered solar and re-absorbed thermal radiation received by the $k$th crater element from the $l$th crater elements on the $i$th facet:
\begin{equation}\label{eq:crscat}
  E^\mathit{crater}_{ik,\mathit{scat}} = E_{\odot}(1-A)(1-s_i) \frac{A}{1-A\frac{\gamma}{\pi}} \frac{1-\cos(\gamma)}{2m} \sum \limits^{m}_{k\neq l} (1-s_k) \cos({z_l})
\end{equation}
and
\begin{equation}
  E^\mathit{crater}_{ik,\mathit{rad}} = \varepsilon_B \sigma_0 (1-A_{th}) \frac{1-\cos(\gamma)}{2m} \sum \limits^{m}_{k\neq l} T^{4}_{l},
\end{equation}
respectively \citep{Emery_etal98}. As in \citet{MacLennan&Emery2019} we use $m=40$ crater elements, which is sufficient resolution for most cases \citep{Spencer90}. The above formulations represent an infinite number of reflections within a crater, as opposed to the single reflection that is assumed between any two facets. If the cratered facet is globally shadowed, then none of the crater elements receive any insolation and $E^\mathit{crater}_\mathit{scat} = 0$. Crater shadowing is also accounted for wherein the scheme presented by \citet{Emery_etal98} is used to determine if the $k$th crater element is shadowed from the Sun by the walls of the crater ($s_k = 1$) or not ($s_k = 0$).

The view factor between a crater element, $l$, on facet $j$ and the $k^{th}$ crater element on facet $i$ is calculated using the view factor between the facets ($f_{j\to i}$) and emission angles of each crater element (i.e., relative to a flat surface) as \citep{Rozitis+Green13}:
\begin{equation}\label{eq:crview}
    f_{jl\to ik} = v_{l,k} f_{j\to i} \frac{\cos(\theta_{k})\cos(\theta_{l})}{\cos(\theta_i)\cos(\theta_j)},
\end{equation}
where $v_{l,k}$ indicates if there is line-of-sight visibility between the two crater elements. To assess $v_{l,k}$ we implement the same procedure as was used to determine $s_k$, but replace the solar incidence angle with the relevant emission angles. All of the elements are forced to have the same area, because of the constructed geometry of the craters, and no area factor is necessary in \autoref{eq:crview}. In order to be consistent with the single-scattering approximation, \autoref{eq:crview} is calculated only one time:
\begin{equation}
    E^\mathit{rough}_{ik,\mathit{scat}} =  \frac{E_{\odot}A(1-A)}{1-A\frac{\gamma}{\pi}} \frac{1-\cos(\gamma)}{2m} \Bigg[ \sum \limits^{m}_{ki\neq li}\cos(z_{i,l}) + \sum \limits^{m}_{kj} f_{jl \to ik}(1-s_{l,j}) \Bigg],
\end{equation}
as opposed to an iterative approach \citep[i.e.][]{Rozitis+Green13}. In sum, the global scattering and self-heating terms in \autoref{eq:energybal} are calculated as $E_\mathit{scat} = E^\mathit{rough}_{ik,\mathit{scat}}$ and $E_\mathit{rad} = E^\mathit{smooth}_{i,\mathit{rad}} + E^\mathit{crater}_{k,\mathit{rad}}$.

\subsection{Reflected Light Removal}

We must estimate and account for any unwanted contribution of reflected light in the observed fluxes before fitting the TPM fluxes to the observations. Observations from IRAS, UKIRT, Spitzer, Akari and the WISE W3 and W4 bands are dominated by infrared emission from Phaethon's surface, and are therefore not contaminated by any reflected light. However, the short wavelength W1 and W2 of the NEOWISE sightings may have a significant reflected light component. The exact amount depends on the light-scattering behavior and the heliocentric distance, as lower surface temperatures will emit less thermal radiation at shorter wavelengths, allowing reflected light to dominate over any thermal emission. We model the reflected light in W1 and W2 in a similar manner to \citet{AliLagoa_etal13} as detailed below.

The amount of reflected light at a given wavelength received at Phaethon's surface is calculated from,
\begin{equation}\label{eq:reflflux}
F_\mathit{\lambda,obs}(\alpha_\odot,\Delta_\mathrm{km},r_\mathrm{au}) = R_{\lambda}\ \frac{F_{\odot_{\lambda}\mathrm{1au}}}{r^2_\mathrm{au}} \Bigg( \frac{1329\ \km \times 10^{-H_V/5}}{2 \Delta_{\km}} \Bigg)^{2} \Phi_\lambda(\alpha_\odot),
\end{equation}
where $F_{\odot_{\lambda}\mathrm{1au}}$ is the solar spectrum at 1~au,  $H_V$ is $V$-band ($0.55\um$) absolute magnitude, $R_\lambda$ is the reflectance relative to $0.55\um$, and $\Phi_\lambda(\alpha_\odot)$ is the photometric phase function. In \autoref{eq:reflflux}, $H_V$ is converted to flux units and is scaled to the desired observer-centric distance. The two-parameter $H, G$ phase function \citep{Bowell_etal89}, $\Phi_\lambda(H_V,G_V,\alpha_\odot)$, is used to calculate the flux at a given solar phase angle. We use $H_V = 14.2\magnitude$ \citep{Devogele_etal20} and a slope parameter of $G_V=0.06$ \citep{Ansdell_etal14}, the former of which is revised from \citet{Tabeshian_etal19}. Because the phase function is reported for 0.55$\um$, we assume that it is the same for W1 and W2. All observed fluxes are scaled to $r_\mathrm{au} = \Delta_\mathrm{au}$ = 1 in order to compare across all of the sightings, which represent many different observing geometries.

A large source of uncertainty in \autoref{eq:reflflux} is the reflectance value ($R_\lambda$). As a B-type, Phaethon's near-infrared spectrum exhibits smaller reflectance values at larger wavelengths over a range of 0.5$-4\um$ \citep{Licandro_etal2007,Kareta_etal18,Takir_etal20}. Specifically, the reflectance value at $3.5\um$ is approximately 70\% of the value at $0.55\um$ \citep{Takir_etal20}. The reflectance spectrum of Phaethon's throughout the visible and near-infrared bears resemblance to carbonaceous chondrites: CK meteorite spectra \citep{Clark_etal10,deLeon_etal12} and heated CI or CM material \citep{Licandro_etal2007,Kareta_etal21}. Because Phaethon doesn't exhibit any absorption features in the $3\um$ region \citep{Takir_etal20} we considered reflectance spectra of dehydrated carbonaceous chondrites presented in \citep{TrigoRodriguez13}. The reflectances of these meteorites in the 3$-6\um$ range do not significantly ($\sim$2\%) vary from a neutral (flat) slope, therefore we claim that an assumption of equal reflectance for W1 and W2 is reasonably justified.

\begin{figure}
    \includegraphics[width=0.5\linewidth]{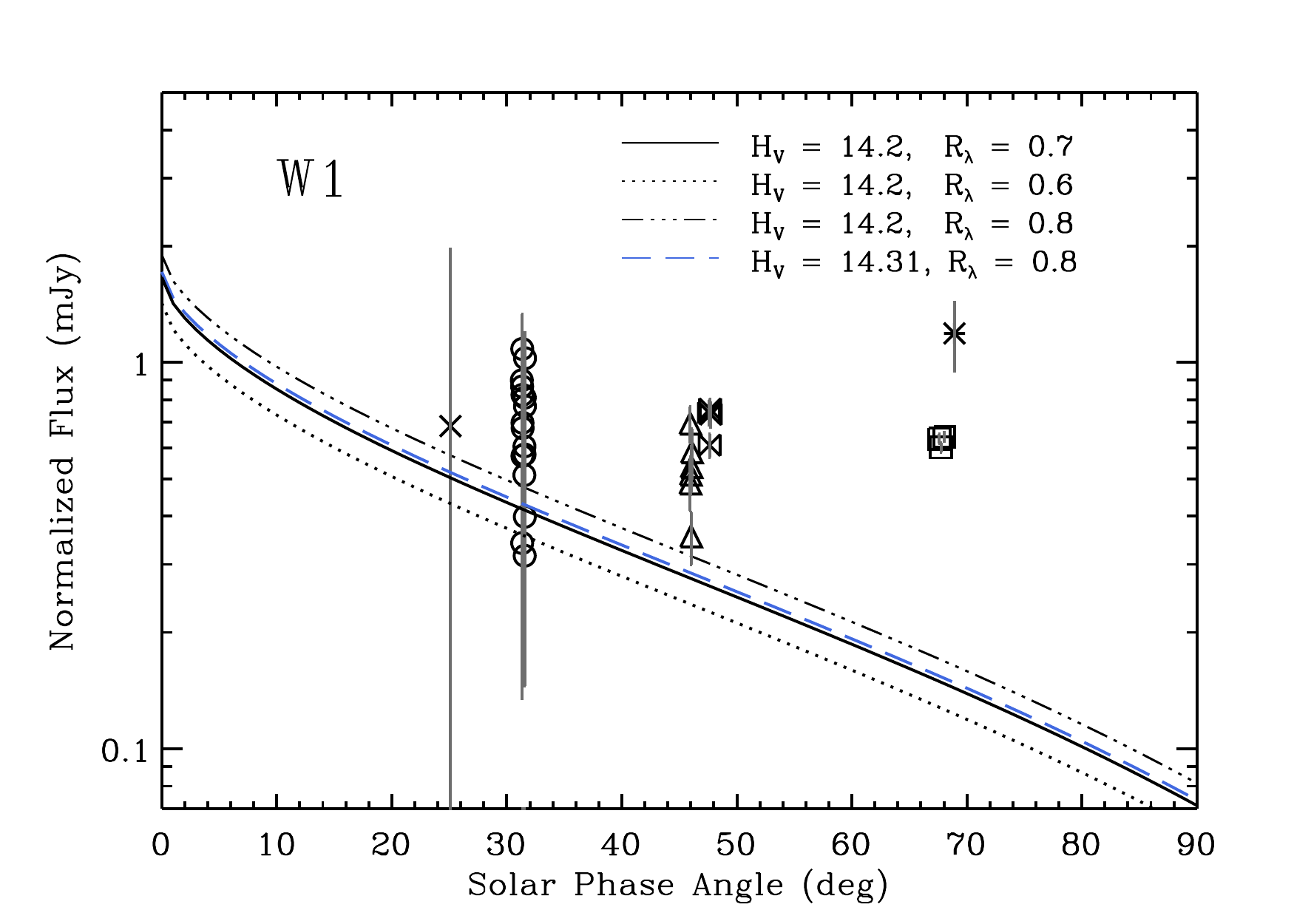}
    \includegraphics[width=0.5\linewidth]{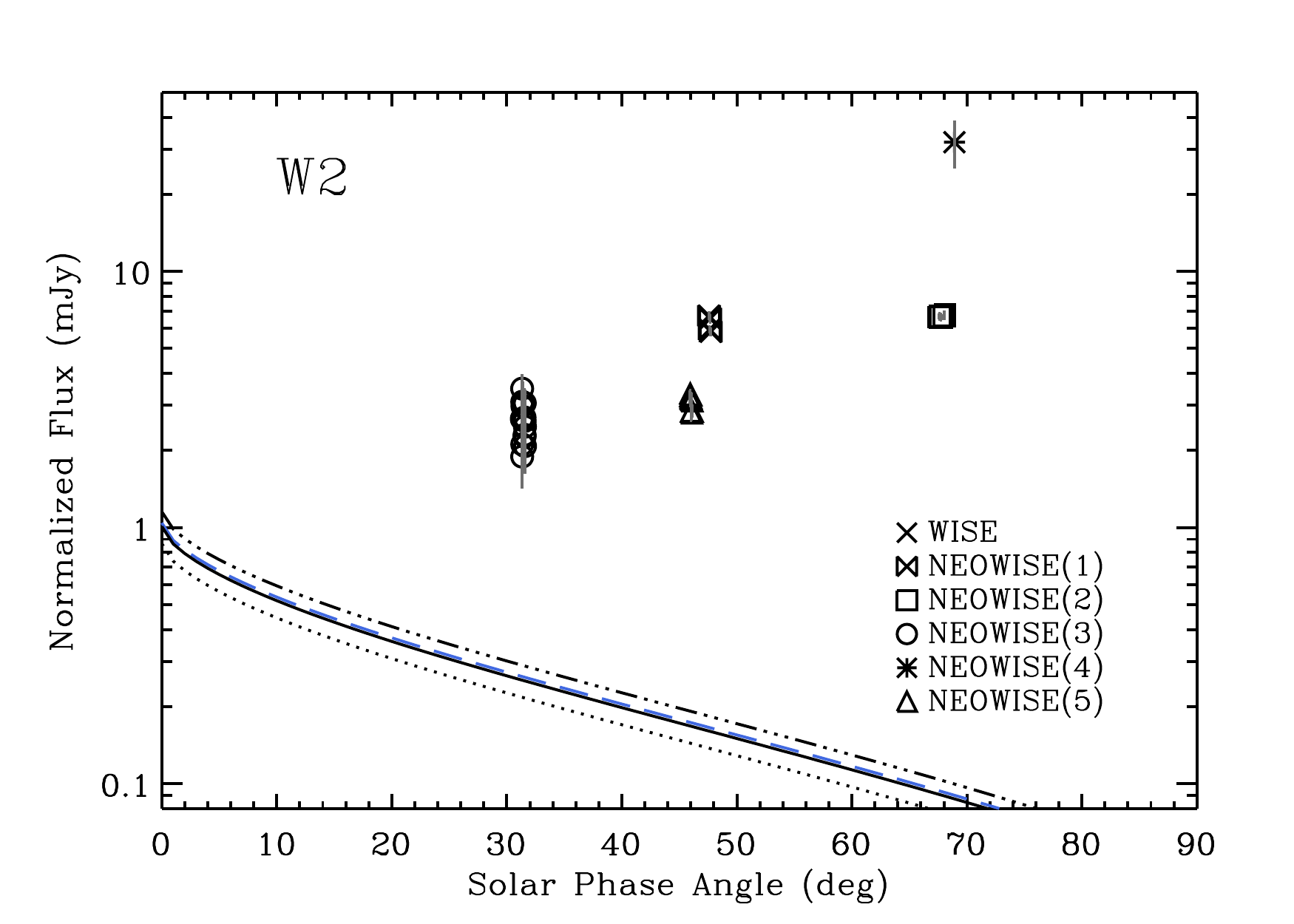}
    \caption{Observed fluxes (black symbols) and uncertainties (grey bars) in NEOWISE W1 (left) and W2 (right) bands, normalized to $r_\mathrm{au} = \Delta_\mathrm{au}$ = 1. The solid and dotted lines are the modeled reflected light phase functions for different reflectivity assumptions (see text for details). The dashed lines encapsulate the uncertainty in the W1 reflectance, $R_{W1}$.}
    \label{fig:refl}
\end{figure}

The observed fluxes in the W1 and W2-bands, along with the modeled sunlight flux are shown in \autoref{fig:refl} where our assumption of $R_{W1,W2} = 0.7$ with $H_V = 14.2\magnitude$ is shown as a solid black line. We also find that an error of 0.1 in $R_{W1,W2}$ is equivalent to an uncertainty of $0.15 \magnitude$, as shown by the dotted and dash-dotted lines. This translates to a relative uncertainty of $\pm 15\%$ in the fraction of reflected light; meaning that a 10\% reflected light component has an uncertainty of $\pm 1.5\%$. Our nominal reflected light model is very close to that found by \citet{Masiero_etal2019} (shown as a blue dashed line in \autoref{fig:refl}), who used $H_V = 14.31\magnitude$ and found $R_\lambda = 0.8$ when using a model that fit reflected and thermal components simultaneously.

\begin{table}[h]
\caption{Estimated Fraction of Reflected Light in WISE/NEOWISE Bands}
\label{tab:refl}
\begin{center}
\renewcommand{\arraystretch}{1.2}
\begin{tabular}{l|ccc|cc}
    \hline \vspace{-0.35cm} \\
    Sighting & $r_\mathrm{au}$ & $\Delta_\mathrm{au}$ & $\|\alpha_\odot\|$ & W1 & W2 \vspace{0.2cm} \\ \hline \\[-1,8em] \hline \vspace{-0.3cm} \\
    WISE & 2.317 & 2.077 & $25.1\deg$ & 1.000 & | \\
    NEOWISE(1) & 1.328 & 0.831 & $47.7\deg$ & 0.378 & \cellcolor{red!25}0.026 \\
    NEOWISE(2) & 1.081 & 0.406 & $67.8\deg$ & 0.236 & \cellcolor{red!25}0.013 \\
    NEOWISE(3) & 1.841 & 1.353 & $31.5\deg$ & 0.605 & \cellcolor{red!25}0.092 \\
    NEOWISE(4) & 1.007 & 0.069 & $68.9\deg$ & \cellcolor{red!25}0.120 & $\cancelto{sat}{0.003}$ \\
    NEOWISE(5) & 1.401 & 0.975 & $46.0\deg$ &  0.523 & \cellcolor{red!25}0.054 \\ \hline
    \multicolumn{6}{l}{ {\bf Notes}: Red cells indicate fluxes that are used in the}\\
    \multicolumn{6}{l}{TPM. The crossed out cell indicates a saturated}\\
    \multicolumn{6}{l}{observation.}
\end{tabular}
\end{center}
\end{table}

Due to the uncertainty with modeling reflected light, and associated model sensitivity, we only use observations that have more than an 85\% contribution from thermal emission. In the case of the NEOWISE(4) sighting, both the W1 and W2 observations were dominated by thermal emission, but the W2-band was saturated \citep{Masiero_etal2019} and is not used in thermal model fits. The red cells in \autoref{tab:refl} that are highlighted in red indicate which W1 and W2 observations that are used in thermal modeling fitting described below.

\subsection{Data Fitting}

In the data fitting procedure, several values of the input parameters are sampled: effective diameter, thermal inertia, and surface roughness. We determine the best-fit effective diameter by searching across all observations, while the thermal inertia is estimated independently for each epoch. For the sightings with a reflected light component, we multiply the observed fluxes by the modeled relative fraction of thermal emission. For each combination of input parameters the chi-squared statistic is calculated, $\chi^2 = \sum (F_{obs,i}-F_{mod,i})^2/\sigma_{obs,i}^2$, where $\sigma_{obs}$ is the flux uncertainty. For comparison between datasets, we use the reduced chi-squared: $\tilde{\chi^2} = \chi^2/\nu$, where $\nu =$(number of datapoints)$-$(number of model parameters) is the degrees of freedom for the relevant subset of data.

\section{TPM Results \& Analysis}

\subsection{Size and Surface Roughness}

We begin our analysis by first placing constraints on Phaethon's size and surface roughness, as these parameters are the same for all observations. To begin, we individually step through each size/roughness combination for each epoch and select the thermal inertia that yields the best-fit solution (i.e., minimizes chi-squared value). Only the IRAS, UKIRT, Spitzer-IRS, Akari, Spitzer-MIPS and WISE observations provided meaningful, independent constraints on Phaethon's size. These data represent observations collected at different orbital positions (in particular, relative to perihelion passage), that modeling work of \citet{MacLennan_etal21} showed {\it may} have some effect on the surface temperatures of a hypothetical idealized asteroid. However, orbital heating effects are not a significant factor for this dataset because of Phaethon's oblique spin axis and because we are using both pre-$q$ and post-$q$ observations (see also discussion in \autoref{sec:disc}). Both the size and roughness are constrained when accumulating the $\chi^2$ values across all the aforementioned datasets.

\begin{table}[h!]
\caption{TPM effective diameter constraints for different default roughness values}\label{tab:TPMdiam}
\begin{center}
\begin{tabular}[pt]{l|lc|lc|lc|lc}

	\hline \vspace{-0.3cm} \\
	 & \multicolumn{2}{c|}{smooth} & \multicolumn{2}{c|}{low} & \multicolumn{2}{c|}{medium} & \multicolumn{2}{c}{high} \\
        & $D_\mathit{eff}$$^a$ & $\tilde{\chi}^2_{min}$$^b$ & $D_\mathit{eff}$$^a$ & $\tilde{\chi}^2_{min}$$^b$ & $D_\mathit{eff}$$^a$ & $\tilde{\chi}^2_{min}$$^b$ & $D_\mathit{eff}$$^a$ & $\tilde{\chi}^2_{min}$$^b$ \vspace{0.2cm} \\ \hline \\[-1,6em] \hline \vspace{-0.2cm} \\
	UKIRT & 4.8 $\pm$ 0.2 & 0.6 & 5.3 $\pm$ 0.2 & 0.6 & 5.3 $\pm$ 0.3 & 1.3 & 5.5 $\pm$ 0.3 & 1.6\\
	Spitzer-IRS & 4.9 $\pm$ 0.5 & 0.3 & 5.5 $\pm$ 0.5 & 0.2 & 6.1 $\pm$ 0.5 & 0.3 & 6.5 $\pm$ 0.5 & 0.3\\
    \multicolumn{1}{r|}{\it pre-$q$} & 4.8 $\pm$ 0.2 & 1.0 & 5.3 $\pm$ 0.2 & 1.2 & 5.4 $\pm$ 0.2 & 4.0 & 6.2 $\pm$ 0.3 & 30.8 \\
    \hline
	IRAS & 6.0 $\pm$ 0.2 & 0.9 & 6.4 $\pm$ 0.2 & 1.0 & 6.6 $\pm$ 0.2 & 1.2 & 6.8 $\pm$ 0.2 & 1.3\\
	Akari & 5.3 $\pm$ 0.4 & $<0.1$ & 5.9 $\pm$ 0.4 & $<0.1$ & 6.2 $\pm$ 0.4 & $<0.1$ & 6.3 $\pm$ 0.4 & $<0.1$ \\
    \multicolumn{1}{r|}{post-$q$} & 5.5 $\pm$ 0.2 & 1.3 & 6.0 $\pm$ 0.2 & 1.5 & 6.2 $\pm$ 0.2 & 1.8 & 6.3 $\pm$ 0.2 & 1.8 \\
    \hline
    Spitzer-MIPS & 5.4 $\pm$ 0.6 & $<0.1$ & 5.4 $\pm$ 0.6 & $<0.1$ & 5.5 $\pm$ 0.6 & $<0.1$ & 5.5 $\pm$ 0.6 & $<0.1$ \\
    WISE & 5.1 $\pm$ 0.2 & 0.9 & 5.7 $\pm$ 0.2 & 1.0 & 6.0 $\pm$ 0.2 & 1.0 & 6.3 $\pm$ 0.4 & 0.9\vspace{0.1cm} \\
    \multicolumn{1}{r|}{pre-$Q$} & 5.2 $\pm$ 0.5 & 1.1 & 5.6 $\pm$ 0.5 & 1.1 & 5.8 $\pm$ 0.5 & 1.1 & 6.0 $\pm$ 0.5 & 1.1 \\
    \hline
    \multicolumn{1}{r|}{\it all} & 5.0 $\pm$ 0.1 & 6.6 &  {\bf 5.4 $\pm$ 0.1} &  {\bf 6.2} & 5.6 $\pm$ 0.2 & 10.6 & 6.2 $\pm$ 0.3 & 33.0 \\ \hline
    \multicolumn{9}{l}{$^a$Effective diameter, $^b$Minimum (best-fit) reduced chi-squared statistic.}

\end{tabular}
\end{center}
\end{table}

Next, we sum the $\tilde{\chi}^2_{min}$ values for each size and roughness combination taken across all sightings to generate a set of best-fit solutions (lower panel in \autoref{tab:TPMdiam}). A $\tilde{\chi}^2$ threshold criterion is used to establish the upper and lower 1-$\sigma$ uncertainties \citep{Hanus_etal2018,MacLennan&Emery2019}, $\tilde{\chi}^2 < \tilde{\chi}^2_{min} (1+\sqrt{2\nu}/\nu)$. The size and roughness determinations are listed in \autoref{tab:TPMdiam} for each individual sighting, for all the sightings in the table, and when grouped by orbital position (pre-$q$ and post-$q$ as defined in \autoref{tab:thermobs}). For example, we find that the best-fit size and roughness combination for the post-$q$ datatsets (IRAS and Akari) is $5.5 \pm 0.2\km$ for a smooth surface. In our discussion (\autoref{sec:disc}) we explore these results in more detail. For now, we determine that the global best-fit solution is $D_\mathit{eff} = 5.4 \pm 0.1\km$ for low roughness. With these size and roughness constraints, Phaethon's thermal inertia can be independently estimated for each sighting.

\subsection{Thermal Inertia}

To begin our thermal inertia analysis, we sum the $\tilde{\chi}^2$ values for each size and roughness combination across all sightings. By doing this, the size uncertainty in each subset of observations is accounted for. The modeled thermal inertia value that corresponds to the $\tilde{\chi}^2_{min}$ is taken as the best-fit value. Standard uncertainties are then determined using the $\tilde{\chi}^2$ threshold criterion as was done for the size and roughness procedure. Although the global best-fit corresponds to low roughness ($\bar{\theta} = 11\deg$), for comparison we estimate thermal inertia across all default roughness values. All sightings except for the IRAS dataset yielded meaningful thermal inertia constraints. Thermal inertia constraints are shown in \autoref{tab:TPMinertia} for each roughness value along with the estimated blackbody temperature ($T_{bb}$) that is determined by fitting a Planck function to the thermal fluxes.

For airless bodies, the effective thermal conductivity can be expressed as a sum of the radiative ($r_r$) and solid ($k_s$) components \citep{Watson64,Gundlach&Blum13,Sakatani_etal17,Ryan_etal20}. The radiative component of conductivity is temperature dependent ($k_r \propto T^3$), which implies that thermal inertia increases with temperature as $\Gamma \propto T^{3/2}$, and that thermal inertia is dependent on the heliocentric distance \citep[$\Gamma \propto r^{-3/4}_\mathit{au}$][]{Delbo_etal15}. Therefore, the surfaces of high-$e$ NEAs such as Phaethon ($e = 0.889$) are well-suited natural laboratories for observing the temperature dependence of thermal inertia. \citet{Rozitis_etal2018} calculated thermal inertias at different heliocentric distances for three NEAs and found that the thermal inertia of each asteroid increased with decreasing heliocentric distance. They found that the heliocentric dependence of thermal inertia depended on spectral emissivity (or, equivalently, the spectral reflectivity) and that, notably, (1036) Ganymed exhibits a variation that is potentially stronger than expected from theoretical predictions. Using TPM results for over 500 asteroids, \citet{MacLennan+Emery21} estimated the temperature dependence of thermal inertia of $\Gamma \propto T^{2.74 \pm 0.29}$, which is also much stronger than the theoretical expectation.

Phaethon's thermal inertia increases with temperature across all sightings and for each default roughness assumed in the TPM. The variation in thermal inertia with heliocentric distance for low roughness can be captured by the power-law \citep{Rozitis_etal2018}:
$\Gamma = \Gamma_0 r_\mathrm{au}^\zeta$ where $\Gamma_0$ is the thermal inertia at 1~au and $\zeta$ is the variation exponent. The best-fit power law parameters for the entire dataset are $\Gamma_0 = 630^{+80}_{-70} \tiu$ and $\zeta = -2.45^{+0.04}_{-0.05}$. We do not see this as a particularly useful encapsulation of the results because heliocentric distance is only a proxy for temperature, therefore its association to thermal inertia is not direct.

\begin{table}[h!]
\caption{Thermal inertia \& surface blackbody temperature at each sighting (sorted by temperature) for $D_\mathit{eff} = 5.4 \pm 0.1$}\label{tab:TPMinertia}
\begin{center}
\begin{tabular}[pt]{l|c|c|c|c|c}

	\hline \vspace{-0.3cm} \\
	&  & \multicolumn{4}{c}{Thermal Inertia} \vspace{0.1cm} \\
	 & $T_{bb}$ (K) & smooth & low & medium & high \vspace{0.2cm} \\ \hline \\[-1,6em] \hline \vspace{-0.2cm} \\

    WISE & 203$\pm$5 & 145$^{+60}_{-40}$ & 95$^{+30}_{-30}$ & 110$^{+40}_{-30}$ & 110$^{+40}_{-30}$ \\
    Spitzer-MIPS & 204$\pm$5 & 530$^{+100}_{-80}$ & 140$^{+50}_{-40}$ & 110$^{+40}_{-30}$ & 90$^{+30}_{-20}$ \\
    NEOWISE(3) & 237$\pm$5 & 135$^{+35}_{-35}$ & 125$^{+30}_{-30}$ & 180$^{+40}_{-40}$ & 200$^{+50}_{-50}$ \\
    NEOWISE(5) & 251$\pm$5 & 400$^{+60}_{-50}$ & 270$^{+50}_{-50}$ & 300$^{+40}_{-40}$ & 225$^{+30}_{-30}$ \\
	Spitzer-IRS & 253$\pm$3 & 1200$^{+800}_{-600}$ & 480$^{+80}_{-70}$ & 380$^{+50}_{-70}$ & 175$^{+30}_{-30}$ \\
	Akari & 266$\pm$8 & 400$^{+140}_{-90}$ & 500$^{+130}_{-90}$ & 600$^{+150}_{-100}$ & 500$^{+110}_{-90}$ \\
	NEOWISE(1) & 269$\pm$5 & 500$^{+80}_{-80}$ & 340$^{+60}_{-60}$ & 380$^{+50}_{-40}$ & 325$^{+40}_{-40}$ \\
	UKIRT & 275$\pm$8 & 840$^{+190}_{-190}$ & 370$^{+90}_{-70}$ & 400$^{+90}_{-130}$ & 300$^{+50}_{-50}$ \\
	NEOWISE(2) & 282$\pm$5 & 650$^{+80}_{-90}$ & 570$^{+90}_{-90}$ & 940$^{+110}_{-120}$ & 1000$^{+180}_{-160}$ \\
	NEOWISE(4) & 313$\pm$5 & 680$^{+700}_{-230}$ & 420$^{+220}_{-180}$ & 500$^{+220}_{-190}$ & 370$^{+200}_{-140}$ \\
	
\hline

\end{tabular}
\end{center}
\end{table}

\begin{figure}
    \centering
    \includegraphics[width=0.7\linewidth]{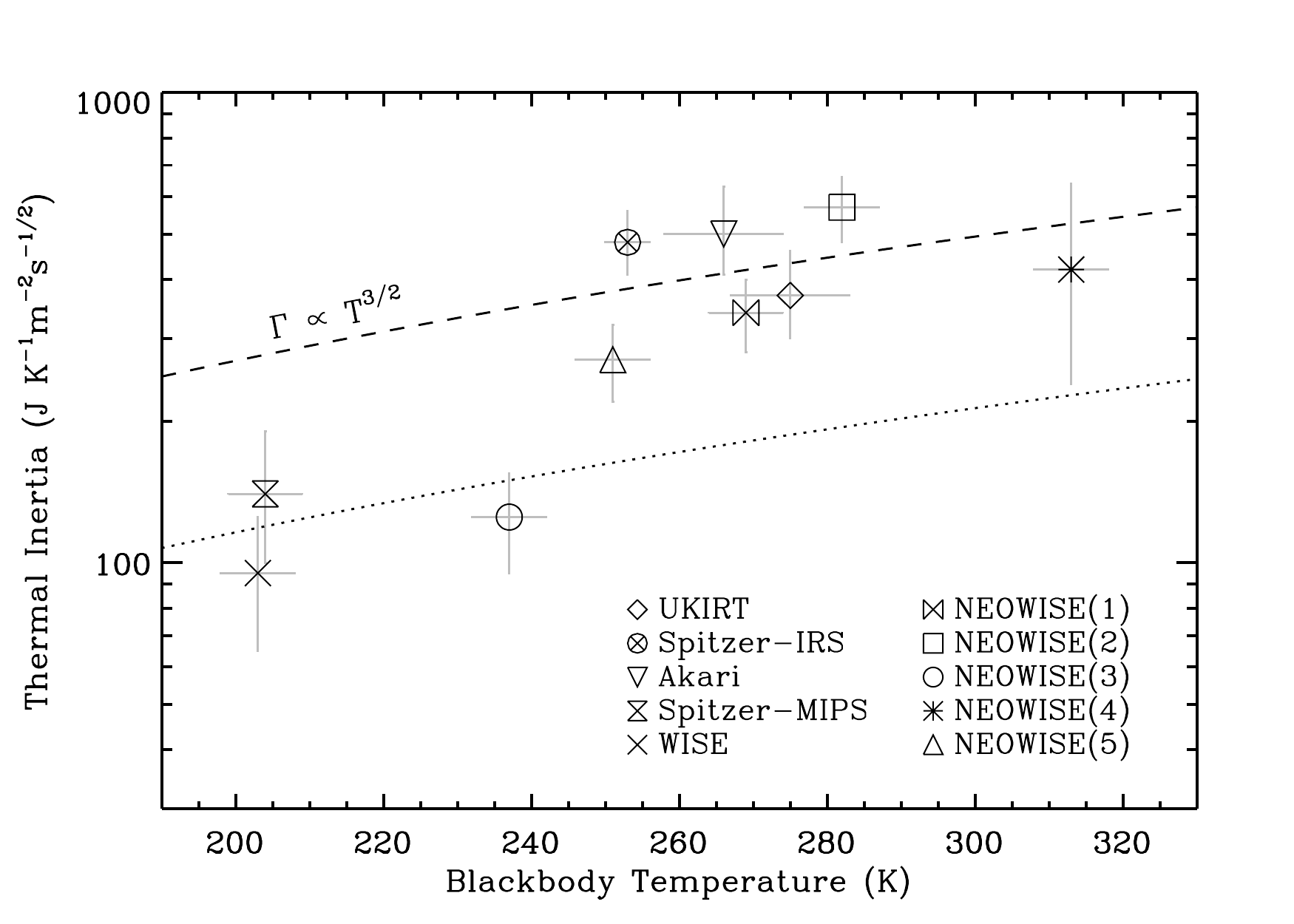} \\
    \includegraphics[width=0.7\linewidth]{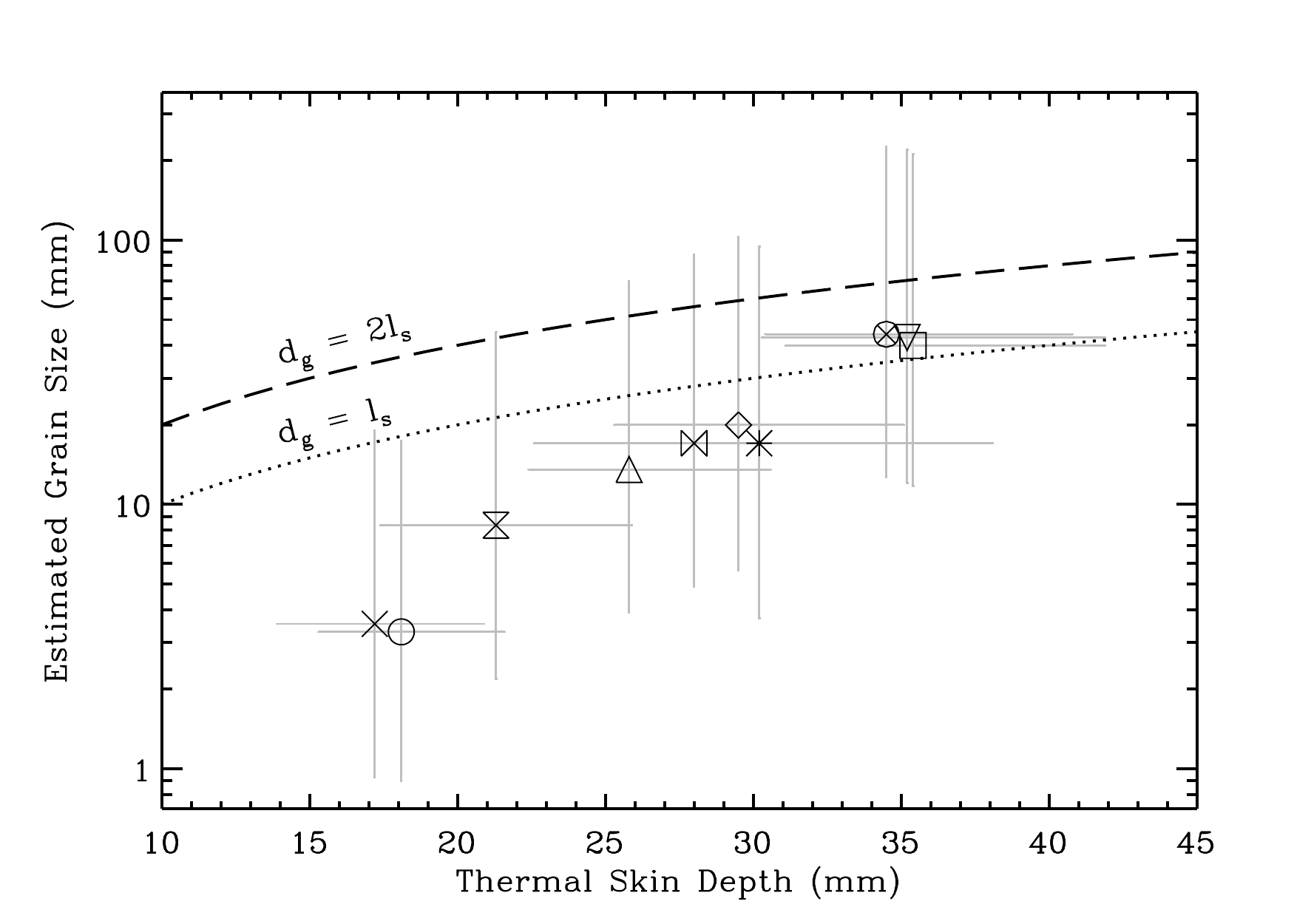}
    \caption{Estimated thermal inertia, $\Gamma$, and blackbody temperature, $T_{bb}$, (top) and, modeled grain diameter, $d_g$, and thermal skin depth, $l_s$, (bottom) for each observational sighting.}
    \label{fig:gsskin}
\end{figure}

Phaethon's thermal inertia variation with temperature (in the range $203\K < T_{bb} < 313\K$) across all datasets can be expressed using the power-law formula: $\Gamma = 1.74\times10^{-9}\ T_{bb}^{4.65}$. Such an extreme temperature dependency is highly physically unlikely, yet we find that it is possible to match the datatset using two curves of $T^{3/2}$ that are scaled to different intercepts (top panel of \autoref{fig:thermalinertia}). This could be an indicator that there are two distinct regions on Phaethon's surface of differing thermal inertia. We consider such a scenario for Phaethon's surface in the following section. In any case, these results indicate that the theoretical relationship between thermal inertia and temperature is supported. The results of \citet{Rozitis_etal2018} and \citet{MacLennan+Emery21}, should be revisited in future investigations.

\section{Surface Models}\label{sub:analysis}

We now construct different surface models aimed at interpreting Phaethon's variation in thermal inertia in a more geologic context. The surface models used herein are an oversimplification but useful to model the grain size and porosity using a temperature dependent heat capacity and accounting for radiative heat transfer within the regolith. This is crucial for comparing thermal inertias estimated at different heliocentric distances. However, a major caveat is that these grain sizes and porosities are likely not accurate in an absolute sense, although they are useful for model comparisons.

\subsection{Homogeneous Surface}

We now interpret the thermal inertia dataset in order to gain insight into Phaethon's surface, specifically the size of regolith particles/rocks and possible heterogeneity. First, we attempt to explain Phaethon's thermal inertia variations with a homogeneous surface consisting of grains of a single size that do not vary with depth. Because heat transport within the regolith occurs via solid transfer (through grains and grain contacts) and radiation (in the voids between grains) we use the thermal conductivity model of \citep{Gundlach&Blum13}. The radiative and solid thermal conductivity components are summed to model the effective thermal conductivity:
\begin{equation}\label{eq:GBcond}
    k_\mathit{eff} = k_r+k_s = k_\mathit{grain} \Bigg[ \frac{18\pi}{4} \frac{(1-v^2)}{E} \frac{\gamma}{d_g} \Bigg]^{-1/3} (f_1\ \exp([f_2 \psi]))\Xi + 4 \varepsilon_B \sigma_0 e_1 f_k \frac{\psi}{1-\psi} d_g,
\end{equation}
where $d_g$ is the grain diameter, which we will refer to simply as the ``grain size''. Here, a regolith grain is a particle with no porosity and therefore only accounts for solid thermal conductivity throughout the material. The grain density ($\rho_\mathit{grain}$) and grain thermal conductivity ($k_\mathit{grain}$ are respectively taken as $2800 \kg \meter^{-3}$ \citep{Macke_etal19}, and $1.5\W\meter^{-1}\K^{-1}$ \citep{Opeil_etal10}. Poisson's ratio ($v$) and Young's modulus ($Y$) are respectively taken from \citep{Flynn_etal18} $v = 0.14$ and $E = 18.9\times10^8 \Pa$. The volume-filling factor, $\psi$, describes the fraction of space occupied by the grains. The non-isothermal factor, $f_k$, is expressed as a function of a dimensionless solid thermal conductivity, $\Lambda_s$, and accounts for thermal gradients across a single particle \citep{vanAntwerpen_etal12,Ryan_etal20}: $f_k = a_1 \tan^{-1}(a_2\Lambda_s^{-a_3}) + a_4$,
where $a_1=-0.568$, $a_2=0.912$, $a_3=0.765$, $a_4=1.035$, and $\Lambda_s = \frac{k_\mathit{grain}}{4d_g\sigma_0T^3}$ \citep{Ryan_etal20}. When the particle is particularly large or if the temperatures are high then $f_k < 1$, and if $\Lambda_s > 25$ then $f_k$ is set to unity. The thermal inertia is computed from the effective thermal conductivity, temperature-dependent heat capacity ($c_s (T)$), grain density, and volume-filling factor:
\begin{equation}\label{eq:ti}
    \Gamma = \sqrt{k_\mathit{eff} \rho_\mathit{grain} c_s \psi}
\end{equation}
where, the heat capacity is modeled as: $c_s (T) = 2.168 \times 10^{-1} + 42.581 \times 10^{-2}\,T + 4.425\times 10^{-2}\,T^2 - 2.060\times10^{-4}\,T^3 + 2.853\times 10^{-7}\,T^4$ \citep{Opeil_etal20}.

We also follow the procedure in \citet{MacLennan+Emery22} to estimate the thermal skin depth, $l_s$:
\begin{equation}
    l_s = \sqrt{\frac{k_\mathit{eff}P_\mathit{rot}}{\pi \rho_\mathit{grain}c_s \psi}} = \sqrt{\frac{P_\mathit{rot}}{\pi}} \frac{\Gamma}{\rho_\mathit{grain}c_s \psi}
\end{equation}
for each thermal inertia. The thermal skin depth is the length scale over which the diurnal temperature range changes by a factor of $e \approx 2.718$ and the thermal inertia represents the thermophysical properties of the regolith over 1-$2l_s$. As discussed in \citep{MacLennan+Emery22}, the grain sizes estimated from thermal inertia may not be physically representative of the surface. Instead they represent a thermally-characteristic grain size that is most related to the thermal conductivity.

The grain size and skin depths for each epoch are depicted in the bottom panel of \autoref{fig:gsskin}. The estimated grain sizes span an order of magnitude and are positively correlated with the skin depth, which varies by over a factor of two. This correlation is most likely due to the fact that both parameters are strongly influenced by the thermal conductivity. The grain-size estimates are statistically equal to the skin depth for largest values and only the smaller grain-size estimates are smaller than the skin depth. These results can be interpreted in two ways: 1) that the thermophysical properties (e.g., thermal conductivity) vary with depth, or 2) the surface is comprised of a wide distribution of grain sizes and boulders.

In the interpretation of thermal inertia, two assumptions about asteroid surfaces are often made: 1) the thermophyscial properties are constant with depth, and 2) the surface coverage is homogeneous. In what follows, we change these assumptions by first considering a layered two-component regolith model, and then considering a latitude-dependent two-component regolith model. By systematically changing these assumptions, we aim to develop a realistic model that is consistent with Phaethon's thermal inertia estimates.

\subsection{Layered Model}

The layered surface model consists of a regolith over a solid bedrock. The temperature-dependent radiative heat conduction within regolith will cause an increase in the thermal skin depth as the temperature increases. As a result, the thermal inertia (which represents the bulk thermophyscial properties over $\sim 1-2 l_s$) at smaller heliocentric distances (i.e., higher temperatures) will be larger than a regolith with no bedrock. We test various regolith layer thickness values in order to assess its influence on the effective thermal inertia.

The thermal conductivity of the regolith layer is modeled from \autoref{eq:GBcond}. For the underlying bedrock we assign the laboratory-measured values of the thermophysical properties of CM carbonaceous chondrite meteorites. The thermal conductivity and heat capacity values are the same as in the homogeneous model. The porosity, particle size, and thickness of the regolith layer are varied. We express the thickness of the layer in relation to the skin depth of a pure regolith at $300\K$ ($l_{s_{300\K}}$). Surface temperatures are computed for a spherical object at heliocentric distances from 1 to 2.5~au in 0.1~au increments, as well as for the heliocentric distances of the observations themselves (\autoref{tab:thermobs}).

\begin{figure}
    \centering
    \includegraphics[width=0.7\linewidth]{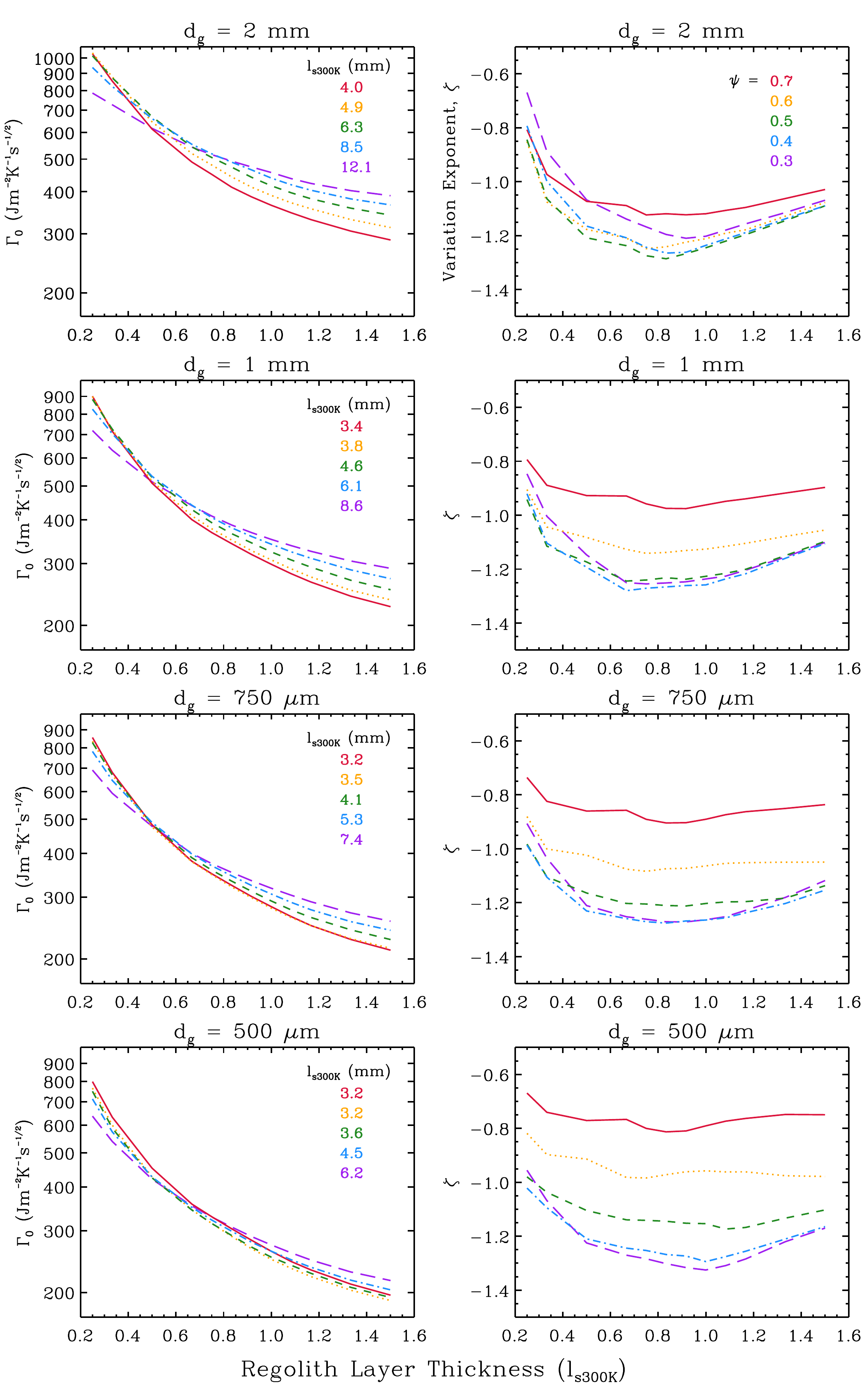}
    \caption{Model results for a layer of regolith comprised of a specified grain size and various porosities overlain atop bedrock. The panels in the left and right columns show the thermal inertia at 1~au, $\Gamma_0$, and the heliocentric variation exponent, $\zeta$, respectively. All values of $\zeta$ are much smaller than that of Phaethon ($\zeta = -2.45^{+0.04}_{-0.05}$).}\label{fig:layered}
\end{figure}

To compare this model to the set of estimated thermal inertias we first calculate the bulk (effective) thermal inertia. This is done by comparing the diurnal temperature variation of the layered model to that of a model that assumes a homogeneous regolith structure. This approach was one of the approaches taken by \citet{Biele_etal19} to model the effective thermal inertia of a bedrock covered by a layer of regolith. We then fit a power law function, $\Gamma = \Gamma_0 r_\mathrm{au}^\zeta$, used in \citep{Rozitis_etal2018} to characterize the variation in thermal inertia with heliocentric distance for all scenarios of the layered surface model. We compute $\Gamma_0$ and $\zeta$ for each combination of the three input parameters: layer thickness, grain size and porosity of the regolith layer, and are depicted in \autoref{fig:layered}.

\subsection{Quasi-Hemispherical Model}\label{sub:quasi}

\begin{figure}
     \centering
      \begin{subfigure}[b]{0.24\textwidth}
         \centering
         \includegraphics[height=\linewidth]{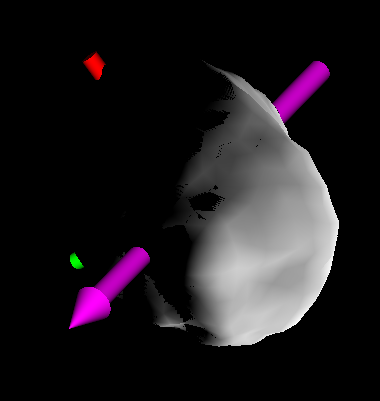}
         \caption{1983-10-11, IRAS}
     \end{subfigure}
      \begin{subfigure}[b]{0.24\textwidth}
         \centering
         \includegraphics[height=\linewidth]{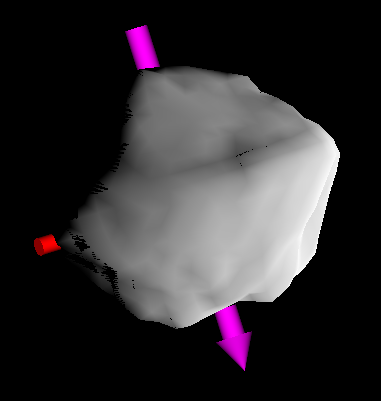}
         \caption{1984-12-20, UKIRT}
     \end{subfigure}
     \begin{subfigure}[b]{0.24\textwidth}
         \centering
         \includegraphics[height=\linewidth]{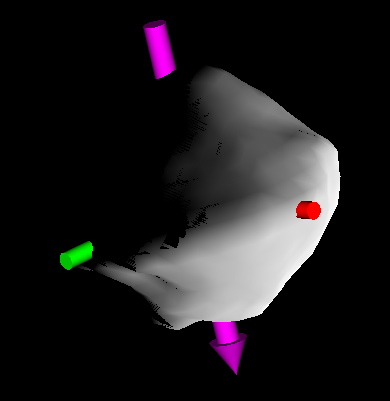}
         \caption{2005-01-14, Spitzer-IRS}
     \end{subfigure}
     \begin{subfigure}[b]{0.24\textwidth}
         \centering
         \includegraphics[height=\linewidth]{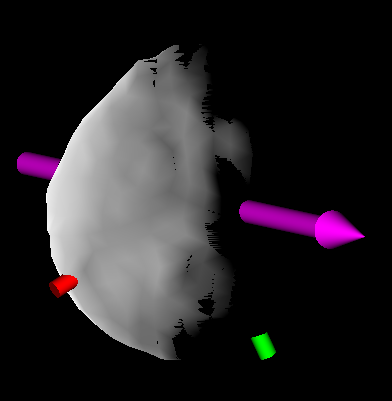}
         \caption{2006-09-23, Akari}
     \end{subfigure} \\
     \vspace{0.3cm}
     \begin{subfigure}[b]{0.24\textwidth}
         \centering
         \includegraphics[height=\linewidth]{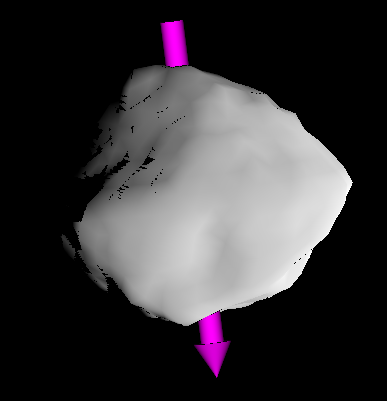}
         \caption{2007-01-01, Spitzer-MIPS}
     \end{subfigure}
     \begin{subfigure}[b]{0.24\textwidth}
         \centering
         \includegraphics[height=\linewidth]{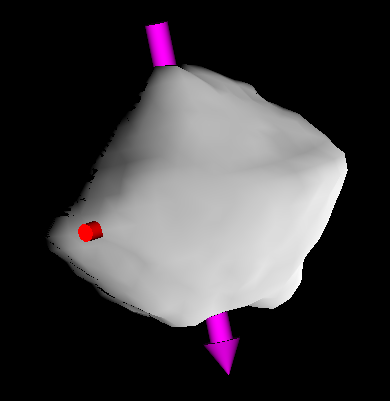}
         \caption{2010-01-07, WISE}
     \end{subfigure}
     \begin{subfigure}[b]{0.24\textwidth}
         \centering
         \includegraphics[height=\linewidth]{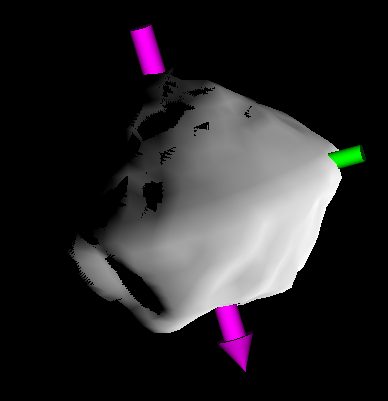}
         \caption{2015-01-13, NEOWISE(1)}
     \end{subfigure}
     \begin{subfigure}[t]{0.24\textwidth} 
        \centering
        \includegraphics[height=\linewidth]{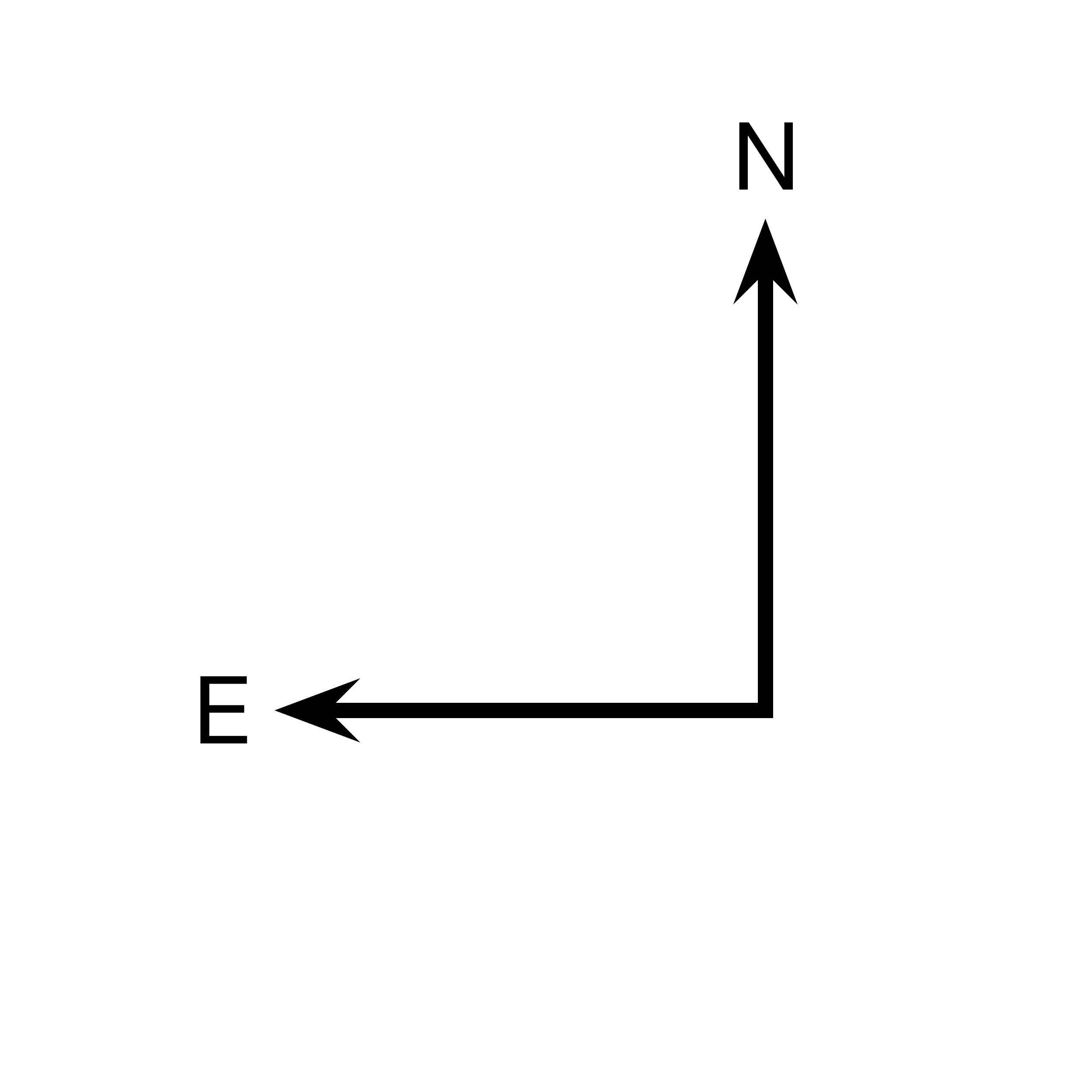}
     \end{subfigure}\\
    \vspace{0.3cm}
     \begin{subfigure}[b]{0.24\textwidth}
         \centering
         \includegraphics[height=\linewidth]{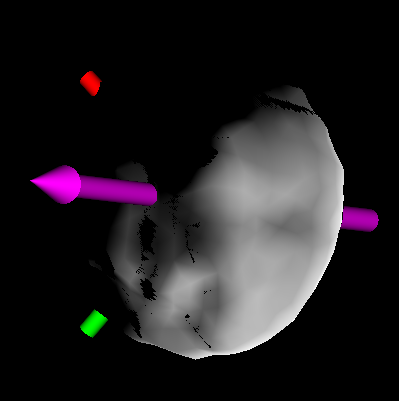}
         \caption{2016-10-02, NEOWISE(2)}
     \end{subfigure}
     \begin{subfigure}[b]{0.24\textwidth}
         \centering
         \includegraphics[height=\linewidth]{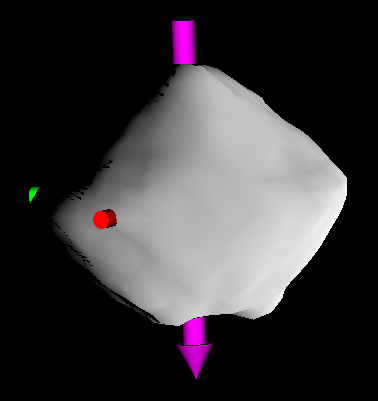}
         \caption{2016-12-06, NEOWISE(3)}
     \end{subfigure}
     \begin{subfigure}[b]{0.24\textwidth}
         \centering
         \includegraphics[height=\linewidth]{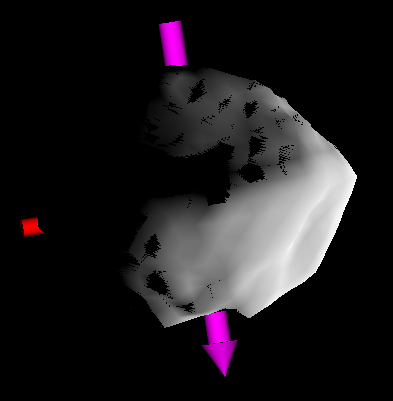}
         \caption{2017-12-17, NEOWISE(4)}
     \end{subfigure}
     \begin{subfigure}[b]{0.24\textwidth}
         \centering
         \includegraphics[height=\linewidth]{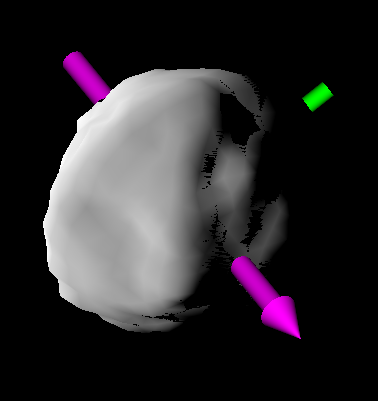}
         \caption{2019-09-07, NEOWISE(5)}
     \end{subfigure}
        \caption{Plane-of-sky views of Phaethon showing the illumination from the Sun for each observing epoch. North is upward and east is leftward in each panel. The magenta arrow is Phaethon's spin axis which points outward from the north pole. The green and red markers appear at $0\deg$ (defined as the long principal axis) and $90\deg$ longitude, respectively. Note that these views are not representative for observation sightings that sample several rotational phases.}
        \label{fig:skyviews}
\end{figure}

We now construct a model that incorporates latitude-dependent grain size, while also accounting for the temperature-dependence of thermal conductivity and heat capacity. In principle, surface heterogeneity can be studied when an asteroid is observed from many viewing aspects. The sub-solar and sub-observer positions indicate the relative amount of flux that is emitted and observed from an asteroid \citep[e.g.,][]{Nugent_etal17}. A greater fraction of thermal flux will be observed from warmer areas (determined by the sub-solar latitude) with smaller emission angles (a function of the sub-observer point). The plane of sky views in \autoref{fig:skyviews} show Phaethon illuminated by reflected sunlight at each observing epoch. Coincidentally, the heliocentric distance of Phaethon correlates with the viewing aspect and therefore the parts of the surface from which the observed thermal flux is emitted. The thermal fluxes acquired at larger heliocentric distances have a significant portion of flux that arises from south of Phaethon's equator, whereas observations in which Phaethon is closer to 1~au mainly sample the northern hemisphere. In addition to the viewing aspect, the wavelength of observation determined the portions of the surface for which thermal fluxes are weighted \citep{Nugent_etal17}. Longer wavelengths are sensitive to cooler surface temperatures which---for a low spin obliquity---are found further from the equator. On the other hand, short wavelength observations sample the warmer equatorial region.

\begin{figure}
    \centering
    \includegraphics[width=0.8\linewidth]{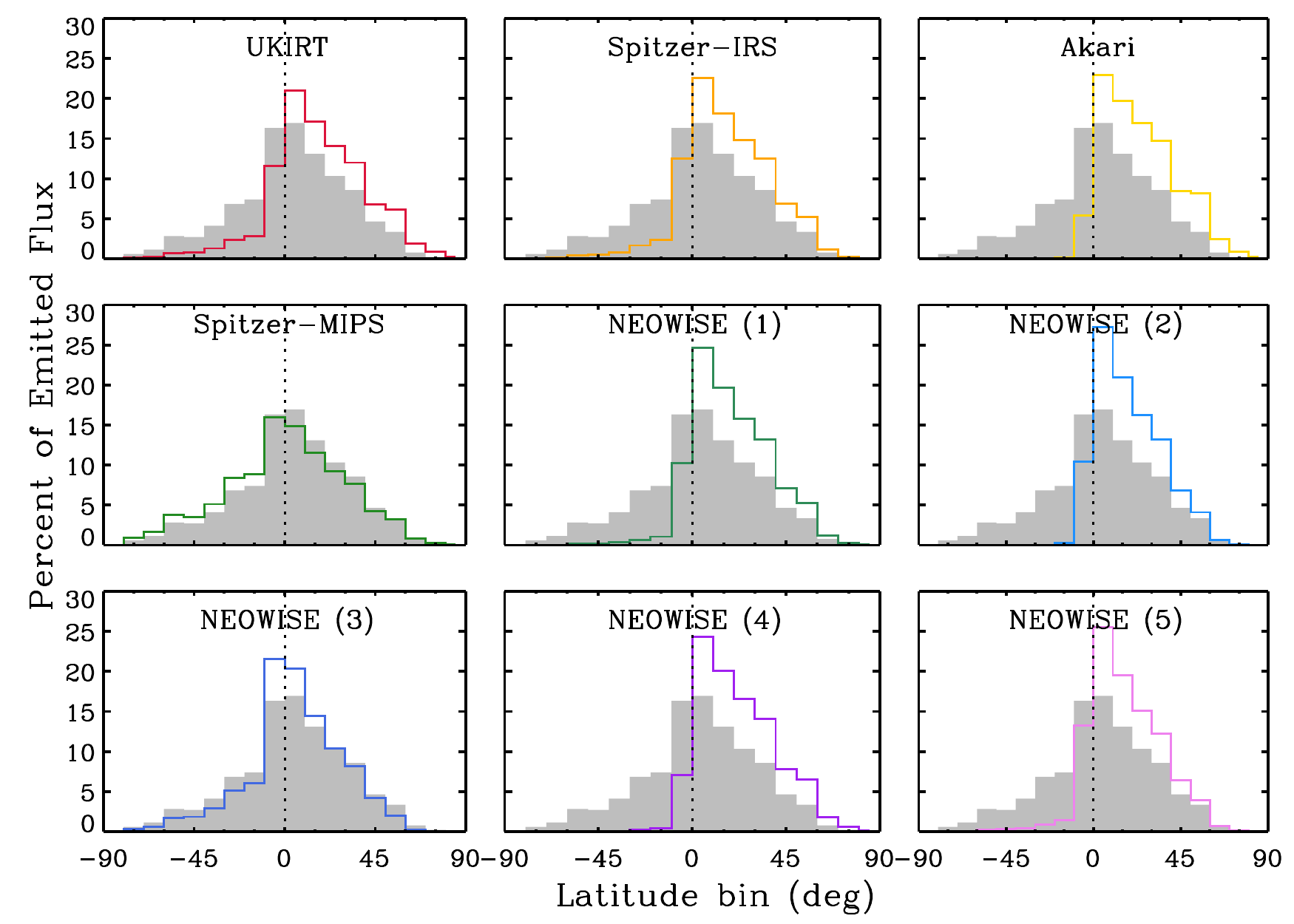}
    \caption{Latitudinal distribution of observed thermal flux from Phaethon's surface for each epoch of observation. The 2010-Jan-07 WISE sighting is shown as the filled grey histogram in each panel for comparison purposes.}
    \label{fig:fluxdist}
\end{figure}

Informed by these principles, we calculate distributions of observed thermal flux emitted from Phaethon's surface for each observing epoch. The emitted flux from the fitting procedure is used to produce rotationally-averaged flux within $10\deg$ latitude bins. The observed thermal flux distributions for each of the ten epochs are shown in \autoref{fig:fluxdist}, and epochs on which Phaethon was observed using multiple filters or a spectrometer are presented as averages of the relevant wavelengths. For comparison purposes, each panel shows the distribution for the WISE 2010-Jan-7 epoch as a filled grey histogram. It can be seen that the NEOWISE(3) epoch differs from the WISE epoch with more flux observed at the equator because of the shorter wavelengths that were active and detected Phaethon. On the other hand, Phaethon's northern hemisphere comprise more than 90\% of the emitted flux observed by Akari and the NEOWISE(4) sighting.

We showed from the thermal conductivity model that smaller grains are a better fit at larger heliocentric distances, and larger grains are more consistent with the smaller heliocentric distances. Thus, we adopt a two-component model by which a larger grain size is assigned to northerly latitudes (ranging from 5 mm to 3 cm) and a smaller grain size for southern latitudes (from 10$\um$ to 400$\um$). Because the upper limit for a modeled grain size is around the thermal skin depth \citep{MacLennan+Emery22} we also consider a boulder-dominated surface for northern latitudes that represents any rock that is greater than $\sim3\cm$. Finally, the porosities of both regions are varied from 10\% to 70\%. The thermal inertia of porous boulders is therefore modeled, but the effect of radiative heat transfer are not included. The boundary latitude between the two regions is varied by 10$\deg$ increments from $-40\deg$S to $+40\deg$N. 
\begin{figure}
    \centering
    \includegraphics[width=0.7\linewidth]{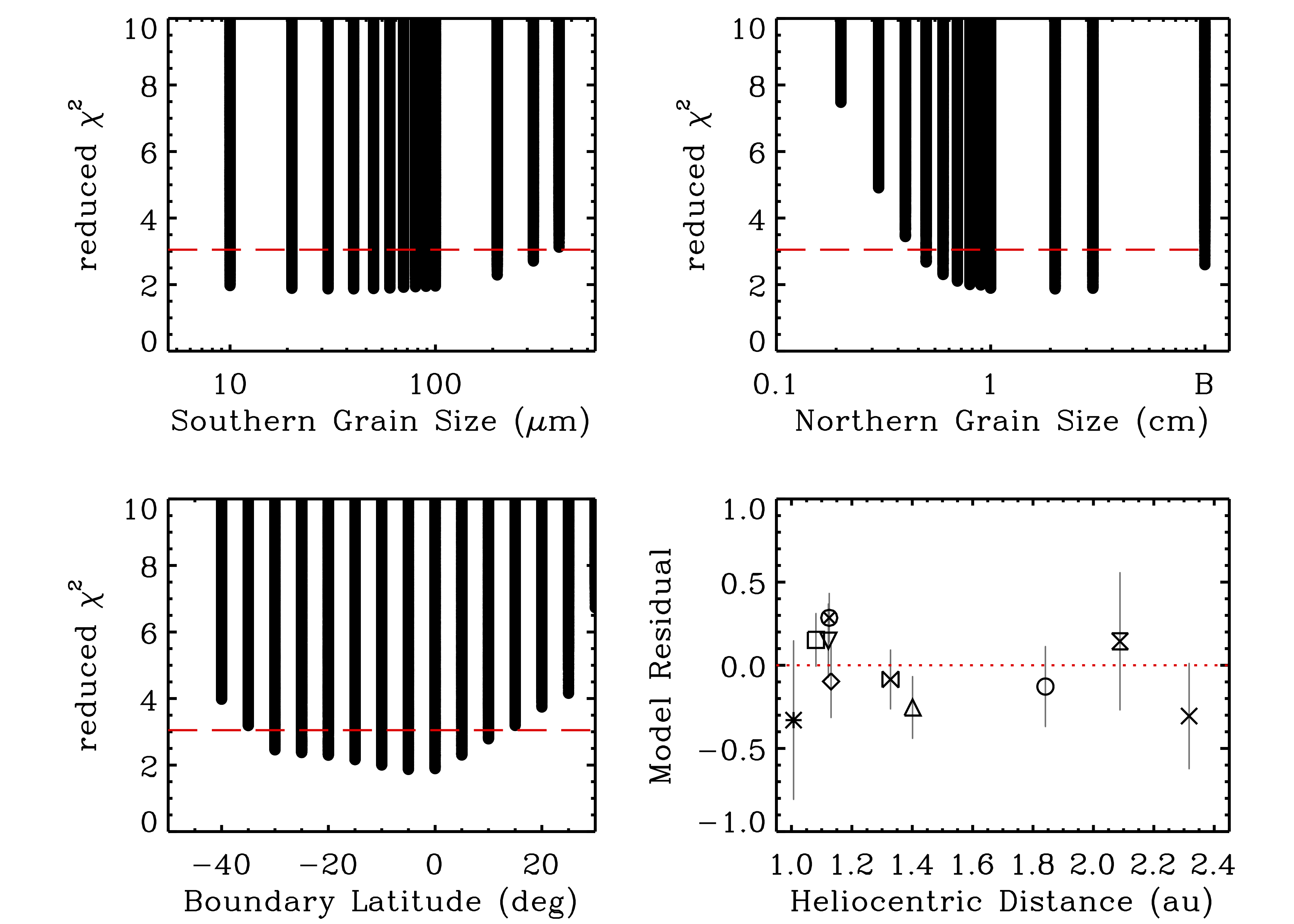}
    \caption{Results of the quasi-hemispherical surface model. Counterclockwise from the top right: grain size fits for the northern latitudes, grain size fits for the southern latitudes, boundary latitude between the two areas, and model thermal inertia residuals. The ``B'' in the upper right panel indicates a bare rock or boulder-dominated surface}. Horizontal red dashed lines indicate the 1$\sigma$ threshold and red dotted line indicates where the model perfectly agrees with the estimated thermal inertia.
    \label{fig:thermalinertia}
\end{figure}

Results of this two-component model are shown in \autoref{fig:thermalinertia}. The best-fit grain sizes (and respective upper and lower limits) for southern and northern areas are 40$\um$ ($<300\um$) and 2$\cm$ ($>5\mm$). A boulder-dominated surface (or bare-rock) surface is represented by a ``B'', indicating widespread presence of grain sizes larger than the thermal skin depth ($3.5\cm$). No lower limit to the grain sizes in the south and no upper limit to the northern region can be determined. This is due to limitations when attempting to model low thermal inertias using \autoref{eq:GBcond} (for southern latitudes), and interpreation issues when the grain size is comparable to the skin depth (for northern latitudes). These grain size constraints can also be expressed in terms of the thermal inertia at 1~au using the previous model inputs with \autoref{eq:GBcond} and \autoref{eq:ti}: $\Gamma_0 = 55^{+60}_{-55} \tiu$ and $\Gamma_0 = 800^{+400}_{-400} \tiu$ for the southern and northern areas, respectively.

\subsection{Equatorial Ridge Model}

Phaethon's spinning top shape consists of a characteristic equatorial ridge---in which the local topography is raised relative to a spherical shape. Ridges such as these can form out of the downward flow of material from higher latitudes because the local gravitational potential is lower at the equator. Therefore we test a scenario in which smaller particles are concentrated in an equatorial band and larger particles or boulders are found at higher latitudes. The width of this equatorial band is varied from $5\deg$ to $25\deg$ in $5\deg$ increments and several particle sizes are tested: 50, 60 70, 80, 90 100, 200, and 300$\um$. Larger particle sizes are assumed for the areas outside the equatorial band 0.8, 0.9, 1, 2, and 3$\cm$. Additionally, we use the thermal inertia of solid CM material to approximate a boulder covered surface. We found that the best-fit combination of parameters result in a $\chi^2_\mathit{min}$ of $\sim10$ that does not indicate any improvement over the $\chi^2_\mathit{min} \sim 1.9$ value found for the quasi-hemispherical model.

\section{Discussion}\label{sec:disc}

\subsection{Size Reconciliation}

Using the radar shape model and combining data from five telescopes we are able to place an estimate on Phaethon's size that is consistent with the radar observations \citep[i.e.,][]{Taylor_etal2019}. We also quickly note here that the best-fit roughness parameters determined herein ($\gamma = 45\deg$, $f_c = 0.5$) are nearly identical to that found by \citet{Hanus_etal2016} ($\gamma = 50\deg$ and $f_c = 0.5$). The size constraints for each data subset listed in \autoref{tab:TPMdiam} present some interesting trends. Our best-fit effective diameter measurements that are based only on IRAS, UKIRT, and Spitzer-IRS observations are $6.4\pm0.2\km$, $5.4\pm0.2\km$, and $5.5\pm0.1\km$, respectively. These values are all systematically larger than, but partially consistent with, the \citet{Hanus_etal2016,Hanus_etal2018} estimates based on the same respective data sets: $6.0^{+0.5}_{-0.3}\km$, $4.6^{+0.4}_{-0.2}\km$, and $5.1\pm0.2\km$. Furthermore, a spherical shape assumption was used to estimate effective diameters of 4.17$\pm0.13\km$ and 4.6$^{+0.2}_{-0.3}\km$ using Akari data and WISE/NEOWISE data, respectively \citet{Usui_etal11,Masiero_etal2019}. Both these values are markedly inconsistent with our estimations of $5.9\pm0.4\km$ and $5.7\pm0.2\km$, for Akari and WISE data, respectively. These results strongly suggests that differences in the shape models used in each work are the cause of the discrepancies in size determination.

\begin{figure}
    \centering
    \includegraphics[width=0.7\linewidth]{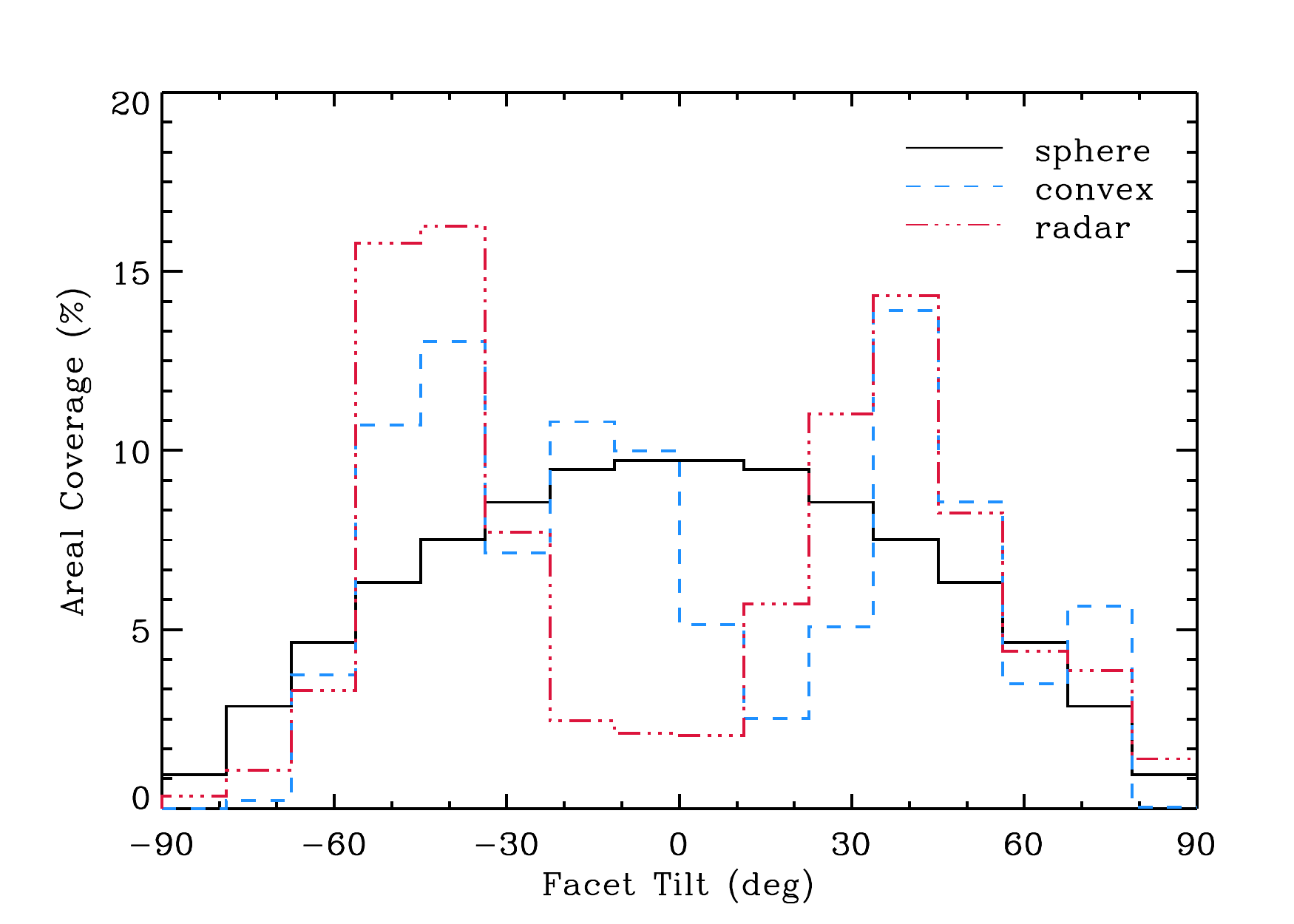}
    \caption{Angle of facet tilt for convex \citep[dashed blue][]{Hanus_etal2016,Hanus_etal2018} and radar shape (dash-dotted red) compared to a sphere (solid black)}
    \label{fig:shapetilt}
\end{figure}

As noted in \citet{Emery_etal14}, one important factor in shape input for thermophysical modeling is the tilt of facets relative to the direction of the Sun. To compare across the three shapes used in TPM analyses of Phaethon, we depict the distribution of facet tilts relative to the surface normal at the equator on a sphere in \autoref{fig:shapetilt}. A sphere has most of the surface area at the equator and tilted outwards ($0\deg$), whereas the spinning top shapes have higher percent of the surface at $\pm 45\deg$. Interestingly, the radar model used in this work has a bi-modal distribution of facet tilts because of the equatorial ridge. Although the convex shape model exhibits similar peaks, it also has a third peak at around $-15\deg$. We suspect that this is most likely due to the lack of a pronounced equatorial ridge on the convex model.

The uncertainty in the $z$-axis of an asteroid shape, which can differ between convex lightcurve shapes and radar shapes, can affect the best-fit thermophysical model parameters \citep{Rozitis+Green14}. To assess possible complications we compare the shape model dimensions of the convex shape model of \citet{Hanus_etal2016} and radar shape model in this work. Using the maximum and minimum projected areas we calculate the axis ratios of an area equivalent ellipsoid ($a/b$ and $b/c$): $a/b_{radar} = 1.09$, $a/b_{convex} = 1.10$, $b/c_{radar} = 1.22$, and $b/c_{convex} = 1.30$. This indicates that stretching the radar shape by 7\% along the $z$-axis will force the same area-equivalent $b/c$ as the convex shape. This difference between shape models is drastically smaller than the case of (1620) Geographos as studied in \citet{Rozitis+Green14}. Using a radar model with the $z$-axis stretched by 7\% applied to Spitzer-IRS spectrum we find best-fit TPM parameters indicating a slightly smaller size of $D_\mathit{eff} = 5.2 \km$ and somewhat larger thermal inertia of 480$-560\tiu$. Unlike Geographos, the radar observations of Phaethon partly viewed the pole \citep{Taylor_etal2019}, and the shape model used herein is partly based on lightcurve observations. Therefore, we posit that any possible uncertainty in the $z$-axis is insignificant.


Lastly, we note that the {\it relative} differences among the independent diameter estimates from each of the observational sightings are consistently offset from another regardless of what shape model is used. For example, the IRAS size estimates are larger than the value from Spitzer-IRS data both in this work and in \citet{Hanus_etal2016}. We show that diameter estimates using the IRAS and Akari observations, which were acquired post-perihelion, and the WISE sighting before Phaethon reached its aphelion point are all larger than the estimates when using the UKIRT and Spitzer-IRS data, which were acquired before perihelion (\autoref{tab:TPMdiam}). The relative offsets in size estimates are consistent with the theoretical prediction by \citet{MacLennan_etal21} that post-perihelion temperatures are larger than what would be calculated from a diurnal TPM. In this scenario, a period of extreme heating during perihelion passage for a surface with a finite thermal inertia introduces a non-negligible seasonal thermal phase lag. The energy absorbed during a short perihelion passage can thus affect temperatures throughout the orbit; mainly at larger heliocentric distances.

However, a direct comparison to the model predictions of \citet{MacLennan_etal21}, which were made assuming a spherical object with zero spin obliquity, are not appropriate. Specifically, the fact that Phaethon's hemispheres have different heating periods that are roughly symmetric with respect to perihelion means that seasonal effects are not as extreme compared to a zero obliquity case. By using both pre- and post-perihelion observations at small heliocentric distances, our analysis inherently mitigates any possible unaccounted for seasonal temperature (and effective diameter) bias. Nonetheless, future work is necessary to investigate biases in the sizes among NEAs with high orbital eccentricities.

The bulk density of Phaethon was estimated to be $1670 \pm 470 \kg \meter^{-3}$ by \citet{Hanus_etal2018} using a convex shape model with an nominal effective diameter of $5.1\km$ and thermal inertia of $600 \pm 200 \tiu$. Because the diameter and bulk density affect the orbital drift \citep[i.e., the Yarkovsky Effect][]{Hanus_etal2016,Vokrouhlicky_etal15}, in a similar way, we can use the $5.4\km$ effective diameter to scale the previous density estimate to be $1580 \pm 450 \kg \meter^{-3}$. This bulk density is larger than that of (101955) Bennu and (162173) Ryugu ($\sim1190\kg \meter^{-3}$\citep{Lauretta_etal19a,Watanabe_etal19}), which are also classified as B-types. Taking CM chondrites as a compositional analog for Phaethon, this bulk density implies a macroporosity of $\sim56\%$, which is comparable to these two other carbonaceous asteroids \citep{Scheeres_etal19}. Along with this macroporosity estimate, Phaethon's top shape and fast spin rate have implications for its internal structure and formation history in comparison to Ryugu and Bennu \citep[e.g.,][]{Barnouin_etal19} that extend far beyond this work.

\subsection{Surface Heterogeneity and Implications for Activity}

From our model analyses of Phaethon's thermal inertia we have shown that the northern and southern hemispheres have different thermophysical properties. Specifically, we can conclude that the thermal inertia of northern latitudes from $\sim10\deg$ to $+60\deg$ is different than southern latitudes from $-30\deg$ to $-60\deg$. The difference in thermal properties could indicate differences in thermally-characteristic grain size, the abundance of boulders, or some combination of the two. The equatorial region from $\sim-30\deg$ to $10\deg$ may likely be a mixture of the northern and southern terrains. On the other hand, we cannot be certain about the surface properties for polar latitudes (greater than $\pm60\deg$), in each hemisphere because very little thermal emission was observed to come from Phaethon's these latitudes. There could be a significant population of larger boulders that evade detection at southern latitudes, for example, or a possibility of fine-grained regolith at the north pole.

The constraint on the latitude boundary in our quasi-hemispherical model approximately ranges from $-35\deg$ to $+15\deg$. It is probable that the equatorial ridge on Phaethon represents the transition from the southern fine-grained surface to the more rocky northern latitudes. Yet, it is unlikely that the transition exists as a sharp step from one hemisphere to another and may be comprised of an intermediate mixture of material from both hemispheres. It's also interesting to note that the equatorial ridge is not constant in latitude. It is conceivable that material flows that have built up the ridge are distributed unevenly. Evidence of variation in the degree of linear polarization with rotation could indicate this, but the data are not specific enough to make such a conclusion.

From our modeling in this work we cannot rule out a scenario in which Phaethon's surface is {\it both} heterogeneous and layered. For example, the southern hemisphere could consist of a finer-grained layer over a coarse-grained or boulder-like bedrock. Because we do not have observations of southern latitudes at smaller heliocentric distances, it is not possible to test such a model. Alternatively, differences in the temperature-dependence of thermal conductivity in the surface could increase or decrease the perceived thermal inertia differences between the hemispheres. For example, if the temperature dependence were weaker than that predicted by radiative heat transfer mechanism, then the difference(s) in thermophysical properties would need to increase in order to explain the thermal inertia observations presented herein. In the very unlikely case that the temperature dependence of thermal conductivity is a stronger function than theoretically predicted, then the differences would be lessened.

Several activity mechanisms have been proposed explicitly for Phaethon, or are reasonably plausible based on the extreme thermal environment \citep{Jewitt&Li2010}: volatilization of some elements \citep{Springmann_etal19,Masiero_etal21}, thermal fracturing \citep{Molaro_etal20}, meteoroid collisions \citep{Szalay_etal19}, radiation pressure \citep{Bach+Ishiguro21}, electrostatic lofting \citep{Kimura_etal22}, and thermal destruction of minerals \citep{Lisse+Steckloff22}. Despite the fact that Phaethon's hemispheres experience similar heating profiles over an orbit \citep{MacLennan_etal21}, we expect that hypothetical volatilization of elements would be associated with southern latitudes where there are more small particles that are readily liberated from the surface. A higher thermal inertia for Phaethon's northern hemisphere could indicate a high surface coverage of boulders. The thermal fracturing mechanism is most efficient for rocks that are larger than the thermal skin depth \citep{Molaro_etal20} and near the equator \citep{Hamm_etal19}; we expect it to be more relevant in the northern equatorial region that has a high thermal inertia and a larger diurnal temperature range. On the other hand, smaller grains that are likely found in the southern latitudes are easier liberated from an asteroid surface. Instead of thermal fracturing, it's possible that small particles are lost via radiation pressure and/or electrostatic forces at small heliocentric distances \citep[e.g.,][]{Bach+Ishiguro21,Kimura_etal22}. Certain elements and/or minerals are known to be volatile at large temperatures \citep{Springmann_etal19,Masiero_etal21,Lisse+Steckloff22}, and thus could drive dust activity at small heliocentric distances. A volatile mechanism is consistent with the relatively low abundance of sodium found among Geminid meteors \citep{Kasuga_etal06,Abe_etal20}.

The DESTINY$^+$ spacecraft will not, unfortunately, visit Phaethon during its perihelion passage, but instead at 1~au where no dust activity has been detected \citep{Jewitt_etal2019,Ye_etal21}. Hypothesized activity mechanisms can be investigated with the information collected by DESTINY$^+$ Dust Analyzer (DDA) instrument \citep{Arai_etal2018,Arai_etal2019} and images of the surface \citep{Ishibashi_etal22}. The latter could involve identification of recently disturbed sites by mapping Phaethon's surface colors and boulder coverage, and interpretation of morphological features. The DDA will measure the mass, charge, density, and composition of collected particles throughout the entire mission \citep[e.g.,][]{Kruger_etal19}, with the aim of identification of dust that originated from Phaethon. This may be particularly useful for testing the impact-driven activity hypothesis \citep{Szalay_etal19}.

\subsubsection{Spectral Variation}\label{sub:spec}

Spectral slope differences in the UV region were noted by \citep{Licandro_etal2007} to potentially indicate variations in surface properties. Using published spectra and new observations from the 2017 close-Earth approach, \citet{Lee_etal19} found no clear evidence for surface heterogeneity from spectral slope values. Specifically, no outlier observations were identified across many sampled longitudes of Phaethon, but the authors remarked that spectral slopes showed a slight decrease within the range of uncertainty when northern latitudes became more visible to Earth. This scenario is supported by variations in $B-V$ colors, which show a distinct change when the sub-observer latitude is larger than $+20\deg$ \citep{Tabeshian_etal19}. Spectral observations gathered within a few days of \citet{Lee_etal19} by \citet{Lazzarin_etal19} showed a clear decrease in spectral slope from one night to the next. This change was explained to be due to the changing viewing aspect, in which the surface portion above $+70\deg$ disappear from view over the observations, implying that the spectral differences are due to latitudinal variation in surface properties.

Because of the lack of absorption features among B-types, the groups in \citet{deLeon_etal12} were mainly distinguished by spectral slope with Phaethon representing the bluest (most negative) slope. Linking asteroid spectra to meteorite types, however, can be confounded by factors such as grain size which has been shown to affect the spectral slope. A more negative spectral slope is correlated with an increase in grain size and increased thermal alteration \citep{Johnson&Fanale73,Cloutis_etal12}. Laboratory spectra of meteorite chips in these meteorite groups are less red (more blue-sloped) than particulates and thermal alteration decreases the spectral slope \citep{Cloutis_etal12}. Thus, these variations seen in reflectance spectra are consistent with particle size differences. In addition, both red and blue spectral slopes have been measured for CM chondrite Murchison caused by changes in surface roughness \citep{Binzel_etal15}. Finally, the spectrum of ultra-blue (153591) 2001 SN263 resembles that of a chip of the thermally altered CY chondrite Y-82162 \citep{Perna_etal14}. Thus, Phaethon's blue slope may be a result of the large grain size of its regolith, surface roughness, thermal alteration, or any combination of the three. 

\subsubsection{Polarimetric Variation}\label{sub:polar}

Linear polarization ($P_r$) measurements of Phaethon have been gathered and presented in several works to date. Polarimetric observations were gathered in the autumn of 2016 by \citet{Ito_etal18} and throughout Phaethon's 2017 apparition by \citet{Borisov_etal18}, \citet{Devogele_etal18}, \citet{Shinnaka_etal18}, and \citet{Okazaki_etal20}. Differences in the phase polarization curve from these two years suggest some level of heterogeneity \citet{Shinnaka_etal18,Okazaki_etal20}. We summarize the dates in \autoref{tab:polar} and, when possible, extrapolate $P_r$ trends to $\alpha_\odot = 100\deg$. Notably, the 2016 and 2017 polarimetric phase curves are offset at large phase angles by roughly 10-12\%.

\begin{figure}
    \centering
    \includegraphics[width=0.7\linewidth]{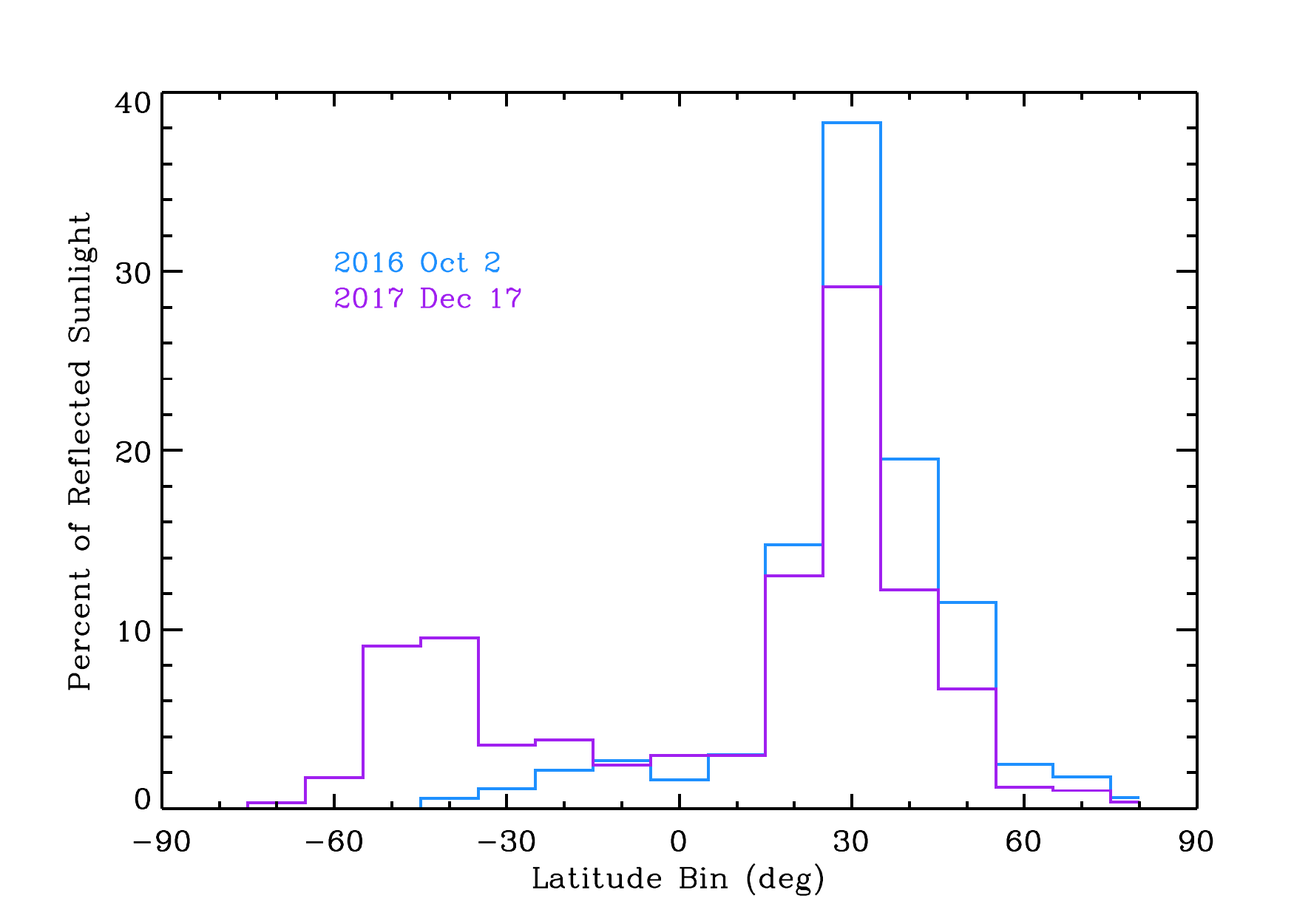}
    \caption{Percent of reflected sunlight as observed from Earth during the approximate dates when polarimetric measurements were acquired (\autoref{tab:polar}).}
    \label{fig:polarsol}
\end{figure}

In order to reconcile these differences in polarimetric properties in the context of our findings we calculate the surface distribution of reflected sunlight that was observed during the 2016 and 2017 sightings. For this exercise we use the same dates as the NEOWISE(2) and NEOWISE(4) observations that approximately match the dates of the $P_r$ measurements of these works. Akin to the thermal flux distribution depicted in \autoref{fig:fluxdist}, we show the latitudinal distribution of reflected sunlight observed from Earth on 2016 Oct 2 and 2017 Dec 17 in \autoref{fig:polarsol}. There is a clear difference in the latitude coverage of reflected sunlight between the two dates, with the 2016 observations highly skewed towards the northern hemisphere and the 2017 observations covering both hemispheres.

The polarimetric phase curve based on 2016 observations is steeper (i.e., a larger $P_r$ variation with changing $\alpha_\odot$) than the curve constructed from 2017 observations. Because higher $P_r$ at larger phase angles are associated with larger particle size, we show that the differences in Phaethon's polarization are consistent with larger grain sizes in the northern hemisphere which are in agreement with the conclusions in this work. It is also interesting to note that rotational variations in $P_r$ are on the order of $\sim 5\%$, which may indicate longitudinal heterogeneity \citep{Borisov_etal18} that was not found in spectrophotometric observations \citep{Lee_etal19}. Searching for longitudinal variation in thermal inertia is technically difficult and beyond the scope of this work, but the residuals from the quasi-hemispherical model fit (\autoref{fig:thermalinertia}) are possibly consistent with this scenario. As we noted earlier, the equatorial ridge may consist of material transported from higher latitudes, making it a likely location for longitudinal changes in surface properties.

\begin{table}[h!]
    \caption{Summary of polarization measurements of Phaethon at large phase angles.}\label{tab:polar}
    \centering
    \begin{tabular}{lcr}
    \hline \vspace{-0.3cm} \\
    UT Date Range & $P_r$ ($\alpha_\odot = 100\deg$)$^a$ & Ref. \\ \vspace{-0.3cm} \\ \hline \\[-1,4em] \hline \vspace{-0.3cm} \\
    2016 Sep 15-Nov 07 & 48\% & \citet{Ito_etal18} \\
    2017 Dec 09-21 & 38\% & \citet{Shinnaka_etal18} \\
    2017 Dec 14-16 & [37\%] & \citet{Okazaki_etal20} \\
    2017 Dec 14-21 & 36\% & \citet{Devogele_etal18} \\
    2017 Dec 15 & [39\%] & \citet{Borisov_etal18} \\
    \hline
    \multicolumn{3}{l}{$^a$ Linear polarization ratio expressed as a percentage.}\\
    \multicolumn{3}{l}{{\bf Note:} Values in brackets are linear extrapolations of} \\
    \multicolumn{3}{l}{observations acquired at smaller phase angles.}
    \end{tabular}
\end{table}

\subsection{Relationship to 2005 UD}

Evidence for an association to (155140) 2005~UD has been suggested on orbital \citet{Ohtsuka_etal2006}, spectroscopic \citet{Licandro_etal2007,Kareta_etal18,Devogele_etal20}, and polarimetric similarities \citep{Devogele_etal18,Devogele_etal20}. Dynamical arguments \citep{Ryabova_etal2019} and observations of dissimilarities in near-infrared spectral slopes \citep{Kareta_etal21} have been used to refute claims that this asteroid pair formed from the splitting of a larger progenitor body in the past. 

The thermal inertia of 2005~UD was determined by \citet{Devogele_etal20} to be 300$^{+120}_{-110} \tiu$ using NEOWISE data at 3.4 and 4.6$\um$. Observations from 2 sightings were included when 2005~UD was at 1.36~au and 1.03~au from the Sun and the thermal emission dominated the W2-band and was significant in the W1-band for the warmer sighting. Because 2005~UD does not have a constrained shape model or spin axis coordinates, it's not currently possible to determine independent thermal inertia for these epochs, as done here for Phaethon. Thus, it is not clear if this reported thermal inertia is more representative of the data collected at 1.36~au or 1.03~au or a weighted combination of both. We suspect that the warmer surface temperatures and better signal-to-noise at 1.03~au may skew the thermal inertia value towards being more representative at this heliocentric distance.

Comparing 2005~UD's thermal inertia to the values derived in this work, one can infer that its surface properties may be a mixture of Phaethon's two components (\autoref{sub:quasi}), which have higher and lower thermal inertias. The grain size estimated from its thermal inertia are in the range of 1-$10\mm$ \citep{Devogele_etal20}. This value lies between the grain sizes found to best represent Phaethon's northern and southern areas. Interestingly, the polarization measurements of 2005~UD \citep[$P_r$ ($\alpha_\odot = 100\deg) \approx 50-55\%$;][]{Ishiguro_etal22} most closely match the 2016 polarimetry (\autoref{tab:polar}), which are more representative of the Phaethon's northerly areas (\autoref{fig:polarsol}). It is possible that 2005~UD formed out of material ejected from Phaethon's northern and southern hemisphere.

\citet{Binzel_etal15} showed that Bennu's near-infrared spectral slope varied from positive and negative values, which they attributed to fine-grained material in the equatorial region. Spectral mapping analyses of OSIRIS-REx observations later supported this claim. Furthermore, our discussion about Phaethon's spectral characteristics mentions grain size and thermal heating effects on the spectral slope. The maximum temperature that 2005~UD currently reaches is just under 1000 K and its perihelion has been larger in the recent past, suggesting that its thermal history is different from Phaethon over the past several thousand years \citep{MacLennan_etal21}. Because the surface roughness of 2005~UD is not currently known \citep{Devogele_etal20}, it could be a yet unaccounted for factor when comparing the near-infrared spectra of the two unusual objects.

\section{Conclusions}

Based on our TPM modeling and analysis of Phaethon's rich thermal infrared dataset we conclude the following:

\begin{enumerate}

\item When incorporating the detailed non-convex shape, Phaethon's size as determined from thermal infrared observations ($D_\mathit{eff} = 5.4 \pm 0.1\km$) is consistent with the radar-derived size. From this effective diameter we revise the bulk density of Phaethon to be $1580 \pm 450 \kg \meter^{-3}$, which is slightly lower than the value reported in \citet{Hanus_etal2016} of $1670 \pm 470 \kg \meter^{-3}$.

\item Phaethon's surface is heterogeneous, with a coarse-grained, or boulder-dominated, northern hemisphere with a higher thermal inertia, compared to southern latitudes dominated by fine grains. The quasi-hemispherical model estimates the north-south boundary latitude to be in the range of -30$\deg$ and +10$\deg$. 

\item The different surface regions described by the quasi-hemispherical surface model (\autoref{sub:quasi}) are consistent with previous evidence from spectral and polarimetric observations (\autoref{sub:spec}, \autoref{sub:polar}) that Phaethon is heterogeneous. Smaller particles that are likely found among southern latitudes are more easily ejected than larger particles in the northern hemisphere.

\end{enumerate}

\section*{Acknowledgements}

EM and, partly, MG were supported by the Academy of Finland through grants 316292 and 299543, respectively. SM was supported by NASA's Near-Earth Object Observations Program through grant 80NSSC19K0523. The Arecibo Observatory is a facility of the National Science Foundation, operated under cooperative agreement by the University of Central Florida in alliance with Yang Enterprises, Inc. and Universidad Ana G. M\'{e}ndez. This publication makes use of data products from the Near-Earth Object Wide-field Infrared Survey Explorer (NEOWISE), which is a joint project of the Jet Propulsion Laboratory/California Institute of Technology and the University of Arizona. NEOWISE is funded by the National Aeronautics and Space Administration (NASA). This research is also partly based on observations with AKARI, a JAXA project with the participation of ESA, and the Spitzer Space Telescope, which was operated by the Jet Propulsion Laboratory, California Institute of Technology under a contract with NASA.

\bibliography{bibfile}

\appendix

\section{Facet View Factors}\label{appA}

Following the work of \citet{Shapiro83}, we compute view factors for each facet pair that are mutually visible. A common and straightforward approach to computing view factors is to integrate across the areas of each facet:
\begin{equation}\label{eq:areaview}
     f_{j \to i} = \frac{1}{2\pi A_j} \oint_{A_i} \oint_{A_j} \frac{(r_{ji} \cdot d\vec{A_i})(r_{ji} \cdot d\vec{A_j})}{||\vec{r^2_{ji}}||}.
\end{equation}

\begin{figure}[h]
\centering
\includegraphics[width=0.4\linewidth]{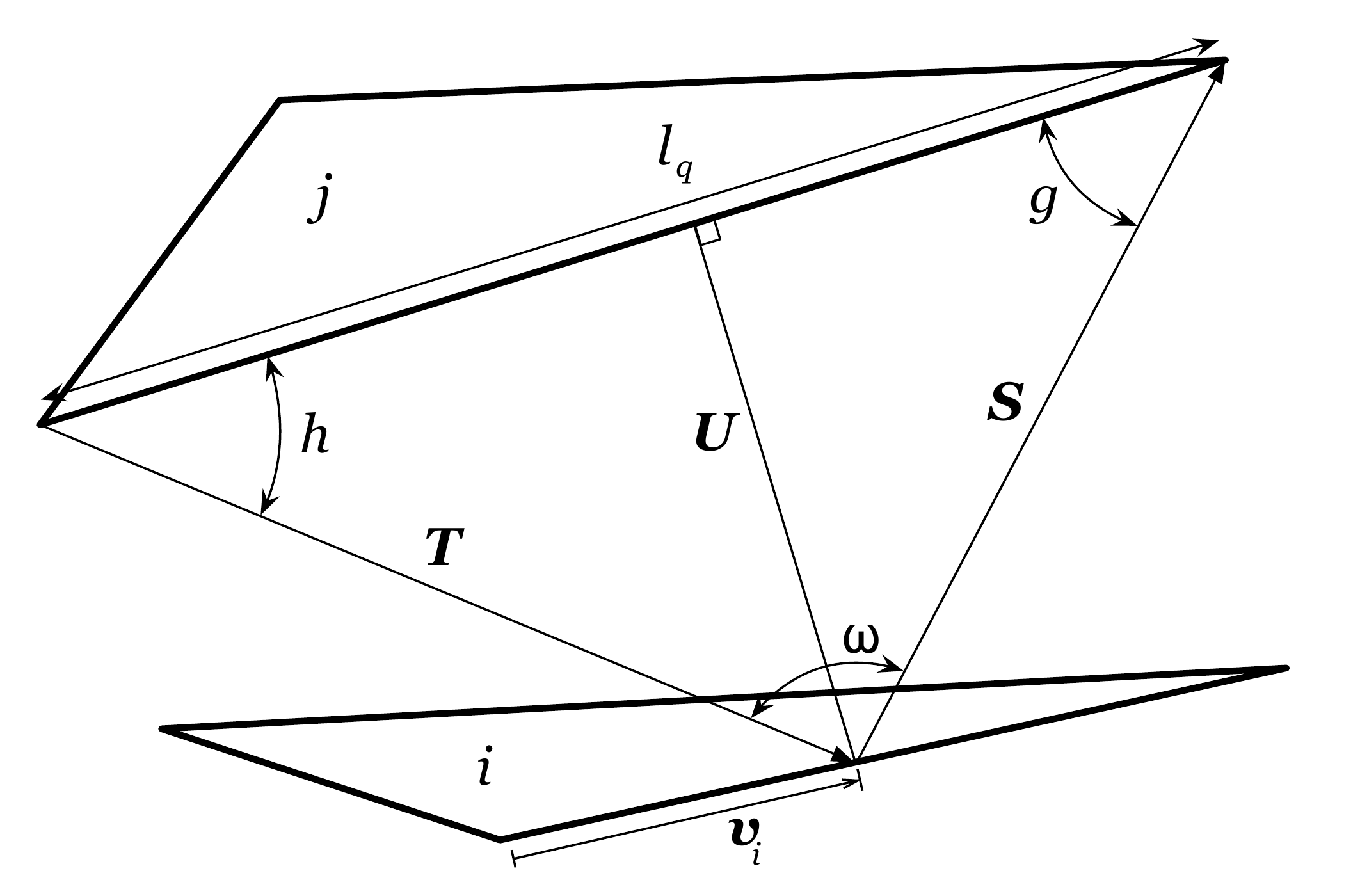}
\caption{Depiction of symbols from \autoref{eq:viewana1} used to compute the view factor between facet $i$ and $j$. Bold symbols indicate vectors rather than scalar terms.}\label{fig:viewfac}
\end{figure}

Stokes's Theorem is used to convert the above area integral into closed line integrals. After this conversion, the view factor in \autoref{eq:areaview} is expressed as a double line contour integral,
\begin{equation}\label{eq:viewint}
    f_{j \to i} = \frac{1}{2\pi A_j} \oint_{C_i} \oint_{C_j} \ln(\vec{r}_{ij})\ d\vec{v}_j \cdot d\vec{v}_i,
\end{equation}
where $r_{ij}$ is a vector connecting the infinitesimal elements $d\vec{v_i}$ and $d\vec{v_j}$ belonging to the edges of their respective facets. In practice, double integral is evaluated around the edges of each facet in a closed loop. One of the integrals in \autoref{eq:viewint} has been solved analytically by \citet{Mitalas+Stephenson66}, which is implemented for two triangular facets as:
\begin{equation}\label{eq:viewana1}
    f_{j \to i} = \frac{1}{2 \pi A_j} \sum^3_{p=1} \sum^3_{q=1} \cos{\Phi_{pq}}\sum^N (\vec{T} \cdot \cos(h) \ln(\vec{T}) + \vec{S} \cdot \cos(g) \ln(\vec{S}) + U \cdot \omega - l_{q})\Delta v_i.
\end{equation}
The vectors $\vec{t}$, $\vec{s}$, and $\vec{v}$, and angles $g$, $h$, and $\omega$ are all functions of $v_i$, as depicted in \autoref{fig:viewfac}, $l_{q}$ is the length of the $qth$ edge on facet $j$, and $\cos{\Phi_{pq}}$ is the cosine of the angle between the $qth$ edge on facet $j$ and the $pth$ edge on facet $i$. With \autoref{eq:viewana1} we divide the $pth$ edge of facet $i$ into $N$ equal segments of length $\Delta v_i$, in order to perform the summation. Repeating this for all edges of facets $i$ and $j$ gives the view factor between them.

Because these formulae require less calculation than the double-area summation method, for a comparable number of subdivisions, it is computationally faster. Furthermore, it has been shown to be more accurate, particularly for adjacent facets, which share a common edge.



Once the view factor from facet $j$ to $i$ ($F_{j \to i}$) is calculated using the procedure outlined above we use the simple reciprocity relationship,
\begin{equation}
    a_j F_{j \to i} = a_i F_{i \to j},
\end{equation}
to calculate the view factor, $F_{i \to j}$, from facet $j$ to facet $i$. This further saves computational time, as the number of view factors that need to be calculated is effectively cut in half.


\end{document}